\documentclass[11pt]{article}
\usepackage{amsmath,amssymb}

\usepackage{epsfig}  
\usepackage{graphicx}               
\usepackage{url}
\usepackage{hyperref}
\usepackage{bm}
\usepackage{slashed}
\usepackage{cite}
\usepackage[dvipsnames]{xcolor}

\usepackage{pstricks}
\usepackage{color}

\newcommand{\nc}{\newcommand}  

\nc{\beq}{\begin{equation}}  
\nc{\eeq}{\end{equation}}  
\nc{\beqa}{\begin{eqnarray}}  
\nc{\eeqa}{\end{eqnarray}}  
\nc{\bit}{\begin{itemize}}  
\nc{\eit}{\end{itemize}}  

\title{  
\vspace*{-2.3cm}
\begin{flushright}
\normalsize{
  }
\end{flushright}
\vspace{1.5cm}
\Large  
\textbf{Phenomenology of Strongly Coupled Chiral Gauge Theories
}\vspace*{1.0cm}   
}

\author{Yang Bai, Joshua Berger, James Osborne and Ben A. Stefanek
\vspace{5mm}
\\
\normalsize\emph{Department of Physics, University of Wisconsin-Madison, Madison, WI 53706, USA}  
}
\date{}

\begin{document}  
\setcounter{page}{0}  
\maketitle  

\vspace*{1cm}  
\begin{abstract} 
A sector with QCD-like strong dynamics is common in models of non-standard physics. Such a model could be accessible in LHC searches if both confinement and big-quarks charged under the confining group are at the TeV scale. Big-quark masses at this scale can be explained if the new fermions are chiral under a new $U(1)^\prime$ gauge symmetry such that their bare masses are related to the $U(1)^\prime$-breaking and new confinement scales. Here we present a study of a minimal GUT-motivated and gauge anomaly-free model with implications for the LHC Run 2 searches. We find that the first signatures of such models could appear as two gauge boson resonances. The chiral nature of the model could be confirmed by observation of a $Z^\prime \gamma$ resonance, where the $Z^\prime$ naturally has a large leptonic branching ratio because of its kinetic mixing with the hypercharge gauge boson.
\end{abstract}  
  
\thispagestyle{empty}  
\newpage  
  
\setcounter{page}{1}  
  
\baselineskip18pt   

\vspace{-2cm}

\section{Introduction}
\label{sec:intro}

One of the simplest examples of physics beyond the Standard Model (BSM) is new fermions in vector-like representations of the SM gauge group. The dynamics of the new fermions is drastically different if they are charged under a new confining non-Abelian gauge group. Condensation spontaneously breaks the chiral symmetry of the fermions and massless Nambu-Goldstone boson (NGB) modes manifest which potentially couple to SM gauge bosons via triangle anomalies~\cite{Hill:2002ap}. The fermions must have an explicitly chiral symmetry breaking mass $m$ in order to lift the NGB's and avoid additional massless degrees of freedom.

If the confinement scale $\Lambda$ is in the range probed by the LHC, the NGB modes are accessible only if $m \lesssim \Lambda$.  If the fermion masses are heavy, $m \gg \Lambda$, the lightest degree of freedom will instead be a spin-zero glueball that is nearly decoupled from the SM even if the fermions are charged under the SM gauge groups. Although technically natural, there is no {\it a priori} reason to expect the fermion masses to be lighter than the confinement scale. This problem is even more acute in models that place the constituent fermions in representations of a Grand Unified Theory (GUT) gauge group. The fermion masses are then generally set by a combination of vector-like mass terms and couplings to the GUT-breaking sector as
\beq
\label{eq:mass-contrib}
- M\,\bar{\psi}\, \psi - y \,\bar{\psi} \,\Phi_{24} \,\psi \,,
\eeq
with $\langle \Phi_{24} \rangle \sim {\cal O}(M_{\rm GUT})$. In order for the resulting fermion masses to be close to but below the confinement scale, these two contributions must either have very small coefficients, or cancel out very finely to arrange a TeV-scale mass. While this is true in GUT models, it can also be the case for example in models of right-handed neutrinos. Technical naturalness protects the fermion masses from large loop corrections, but they can still receive large contributions to their masses when coupled to any singlet which acquires a large vacuum expectation value (VEV).

The SM QCD sector evades this puzzle and indeed has non-decoupled pions close to the scale of strong dynamics. The puzzle is resolved through a combination of factors: small Yukawa couplings, but also a chiral gauge symmetry structure. It is therefore plausible that the new strong dynamics sector operates in a similar way.  Since the fermions cannot be chiral under the GUT gauge group and be consistent with constraints on a chiral fourth generation~\cite{He:2001tp,Novikov:2001md,Kribs:2007nz,Erler:2010sk,Eberhardt:2012gv}, they can only be chiral under a new gauge group.  For simplicity, we study the case where this new gauge group is a $U(1)^\prime$ Abelian group.  If the fermions are charged under this additional gauge group, then the terms of Eq.~\eqref{eq:mass-contrib} are forbidden without additional insertions of a $U(1)^\prime$-breaking VEV.  Given a $U(1)^\prime$-breaking sector near the scale $\Lambda$, the lightest composite particles will also be near $\Lambda$ and will have large couplings to the SM. While there is still a coincidence of scales, it is far less acute. The fermion masses will no longer be near the GUT scale, but rather the scale at which the hierarchy problem is ultimately resolved.

In this paper, we consider a model with an $SU(N_b)$ (big-color) confining gauge group and a $U(1)^\prime$ Abelian group. We also take the fermions in the big-color sector (big-quarks) to transform as fundamentals of an $SU(5)_{\rm GUT}$ symmetry.~\footnote{It is challenging to unify all of $SU(5)_{\rm GUT}$, $SU(N_b)$, and $U(1)^\prime$ together due to anomaly cancellation. Given the SM fermion charges and gauge coupling running, $SU(5)_{\rm GUT}$ unification has additional motivation from experimental data.} Their charge under the Abelian group is chiral, such that the big-quark masses will be forbidden without additional structure. To give the big-quarks mass, we introduce a scalar $\varphi$ charged under $U(1)^\prime$ that couples to the big-quarks. It develops a VEV through its coupling to the confining sector, lifting the big-pion masses to be close to, but below, the mass of $\varphi$. In this way, we construct a chiral model that not only has an experimentally-accessible spectrum, but also that alleviates the tension caused by the coincidence of scales.

The addition of the Abelian gauge group with chiral structure has important and interesting phenomenological consequences.  In particular, there is an additional massive $Z^\prime$ gauge boson coupling to the big-quarks and mixing with the SM $\gamma$ and $Z$. This opens new decay modes for the big-pions, as well as the possibility of direct production of the $Z^\prime$. The chiral structure also leads to additional accidentally approximate discrete and continuous symmetries that lead to three-body decays and potentially long-lived particles.

The remainder of the paper is structured as follows. In Section \ref{sec:model}, we present the detailed structure of the chiral composite model.  Within that section, we study the chiral symmetry breaking pattern in subsection \ref{sec:symmetry-breaking}, the big-pion spectrum in subsection \ref{sec:pion-spectrum}, the properties of the $Z^\prime$ boson in subsection \ref{sec:zprime-interactions}, and the properties of the big-pions in subsection \ref{sec:pion-interaction}. We conclude in Section \ref{sec:conclusion}. In Appendix~\ref{app:real-scalar}, we discuss the perturbativity of models with the elementary real scalar field as a light resonance. We study $U(1)'$ gauge coupling running in Appendix~\ref{sec:gprime-running} and the Yukawa coupling running in Appendix~\ref{app:yd-over-yt}.  Kinetic mixing and decay properties of the $Z^\prime$ are given in Appendix~\ref{app:Zprime-diagonalize}.

\section{The Chiral Composite Pseudoscalar Model}
\label{sec:model}

As described above, we study a model with a confining gauge group $SU(N_b)$ with a confinement scale $\Lambda_b$ at ${\cal O}(\mbox{TeV})$. We introduce big-quarks charged under both the SM gauge group and $SU(N_b)$. Due to the electroweak constraints mentioned above, the big-quarks cannot be chiral under the SM gauge group, so we introduce an Abelian $U(1)^\prime$ gauge symmetry with chiral charges. In order for the $U(1)^\prime$ group to be gauge anomaly-free, we must have two sets of big-quarks. The chiral structure of the charges prevents large quark mass contributions from UV scales such as the GUT scale. As in the SM, we introduce a scalar field $\varphi$ charged under $U(1)^\prime$ that can develop a VEV to spontaneously break $U(1)^\prime$ and give a mass to the corresponding gauge boson $Z^\prime$. We show the content of our model as well as the gauge symmetries in Table~\ref{tab:fieldcontent}.
\begin{table}[ht!]
\renewcommand{\arraystretch}{1.9}
\begin{center}
\begin{tabular}{|cccc|}
\hline \hline
 & $SU(N_b)$    &   $SU(5)_{\rm GUT}$   & $U(1)^\prime$    \\  \hline
$\left(\psi^{\rm T}_{1, L}, \;\psi^{\rm D}_{1, L}\right)$ & $N_b$          &  5                &    $q_1$                       \\ \hline
$\left(\psi^{\rm T}_{1, R}, \;\psi^{\rm D}_{1, R}\right)$  &  $N_b$ &  $5$  &    $q_2$                      \\ \hline  
$\left(\psi^{\rm T}_{2, L}, \;\psi^{\rm D}_{2, L}\right)$ &   $N_b$          &  $\overline{5}$             &    $-q_1$                   \\ \hline 
$\left(\psi^{\rm T}_{2, R}, \;\psi^{\rm D}_{2, R}\right)$  & $N_b$          &   $\overline{5}$            &    $-q_2$                       \\ \hline \hline
$\varphi$     & 1  &  1  &    $q_1-q_2$                       \\ \hline
 \hline
\end{tabular}
\end{center}
\caption{Field content of a chiral model with a confining QCD-like gauge group $SU(N_b)$ and $U(1)^\prime$. Here, $q_1 \neq q_2$. The SM model fermions are neutral under $U(1)^\prime$ and are not listed here. The $U(1)'$ charge assignment shown here to achieve anomaly cancellation is not unique; one may also assign charges of $q_{2}$ and $q_{1}$ for $\psi_{2,L}$ and $\psi_{2,R}$, respectively.
\label{tab:fieldcontent}}
\end{table}
The new fermions, $\psi_{1, 2}$, transform as fundamentals or anti-fundamentals under the $SU(5)_{\rm GUT}$ gauge group. In terms of the SM gauge interactions, $[SU(3)_c, SU(2)_W]_{U(1)_Y}$, we separate them into the QCD color-triplet $\psi^{\rm T}_{1} = (3, 1)_{-1/3}$ and weak-doublet $\psi^{\rm D}_{1} = (1, 2)_{1/2}$, and similarly for $\psi_2$. To have both the $SU(N_b)$ and $SU(3)_c$ gauge couplings asymptotically free in the UV, we require $2 \leq N_b \leq 5$. The case with $N_{b} = 2$ differs from the other allowed values because it has enhanced global symmetry due to the fact that the fundamental representation of $SU(2)$ is pseudo-real.~\footnote{In the case where $N_{b} = 2$, the global symmetry breaking pattern would be $SU(20) \times U(1)_\varphi \rightarrow Sp(20)$ and we would expect (399 - 210) + 1 = 190 PNGB's. The decomposition is $190 =4\times 24 +  3\times(10 + \overline{10}) + 15 + \overline{15} + 1 + 1 + 1 + 1$. Additionally, the $N_b = 2$ case has a perturbative infrared fixed point and is likely to have an approximate conformal symmetry in the IR~\cite{Iwasaki:2003de}.} It also requires the choice of $q_{2}$($q_{1}$) for $\psi_{2,L}$($\psi_{2,R}$) to forbid bare mass terms of the form $\psi^T_{1,L}\,{\cal C}\, \psi_{2,L}$ with ${\cal C}$ as the charge-conjugation operator. The remainder of this paper will focus on the cases with $3 \le N_b \le 5$.~\footnote{There are some debates about whether $N_b = 3$ with $N_f=10$ is inside the conformal window or not~\cite{Ryttov:2007sr,Appelquist:2012nz,Deuzeman:2013kma,Lombardo:2014pda,Fodor:2016zil}. In the later part of our paper, we assume confinement for $N_b=3$ and $N_f=10$.}

The $U(1)^\prime$ charge for the scalar field $\varphi$ is chosen such that it can have renormalizable interactions with the big-quarks in this model. Because of chirality under the additional $U(1)^\prime$ gauge symmetry, there are no bare big-quark masses. On the other hand, the complex scalar field $\varphi$ can have Yukawa couplings to some of new fermions. The allowed renormalizable Yukawa interactions are 
\beqa
{\cal L}_{\rm Yukawa} \supset - y_1^{\rm T} \, \varphi \, \overline{\psi}^{\rm T}_{1, L} \, \psi_{1, R}^{\rm T} \,-  y_1^{\rm D} \, \varphi \, \overline{\psi}^{\rm D}_{1, L} \, \psi_{1, R}^{\rm D} \,- \, y_2^{\rm T} \, \varphi^{*} \,  \overline{\psi}^{\rm T}_{2, L} \, \psi^{\rm T}_{2, R} - \, y_2^{\rm D} \, \varphi^{*} \,  \overline{\psi}^{\rm D}_{2, L} \, \psi^{\rm D}_{2, R}  +  \, h.c.
\label{eq:Yukawa-coupling}
\eeqa
For simplicity, we choose identical, real Yukawa couplings such that $y_1^{\rm T} = y_2^{\rm T} = y_{\rm T}$ and $y_1^{\rm D} = y_2^{\rm D} = y_{\rm D}$. The most general renormalizable potential for $\varphi$ is
\beqa
V(\varphi) = m_{\varphi}^{2} \, \varphi^{*} \varphi \,  + \, \lambda_{\varphi}\, (\varphi^{*} \varphi)^{2} \, +\, \lambda_{\varphi h}\, \varphi^{*} \varphi\, H^\dagger H\,.
\eeqa
with $H$ the Higgs doublet in the SM. The Yukawa coupling of $\lambda_{\varphi h}$ can potentially modify the SM Higgs boson properties by introducing additional decay channels. In light of the good agreement of the Higgs boson properties with the SM, there could be a stringent constraint on $\lambda_{\varphi h}$.  Absorbing the electroweak VEV correction, we define a new mass for the $\varphi$ field as $\overline{m}_\varphi^2 = m_\varphi^2 + \lambda_{\varphi h}\, v_{\rm EW}^2/2$ with $v_{\rm EW} = 246$~GeV. 

If $m_{\varphi}^{2} < 0$, similar to the Higgs field in the SM, the scalar field can develop a non-zero VEV independent of the strong dynamics sector. For $\lambda_{\varphi h}=0$, the VEV is $\langle \varphi\rangle = (- m_{\varphi}^{2} / \lambda_{\varphi})^{1/2}/\sqrt{2}$, which could be far above the TeV scale. For this case, we need to have small Yukawa couplings, $y_{\rm T, D}$, to have small big-pion masses. The situation is very similar to the light flavors in the SM QCD sector, except that for the new $Z^\prime$ gauge boson to be within the low energy spectrum below around 1 TeV its gauge coupling should be small. However, if $m_{\varphi}^2 > 0$ the situation is even more interesting. Because the big-quark condensate can generate a tadpole term for $\varphi$ through the Yukawa interactions, $\varphi$ can still develop a VEV, which is triggered by the $SU(N_b)$ confinement scale $\Lambda_b$. As a result, the $Z^\prime$ gauge boson mass should be related to and likely below $\Lambda_b$. Some big-pions could decay into this $Z^\prime$, which could be a smoking gun to test our model. In our paper, we will mainly concentrate on the case with $m_{\varphi}^{2} > 0$.

\subsection{Symmetry Breaking and Counting PNGB's}
\label{sec:symmetry-breaking}
At the confinement scale of $\Lambda_b = {\cal O}(\mbox{TeV})$, the gauge coupling of $SU(N_b)$ becomes large such that the bi-fermion operator develops a nonzero VEV. For a weak $U(1)^\prime$ gauge interaction~\footnote{The gauge coupling $g^\prime$ should be smaller than 0.35(0.28) for $N_b=3(5)$ to have its Landau pole below the GUT scale; see Appendix~\ref{sec:gprime-running}.}, following the vacuum alignment argument in Ref.~\cite{Peskin:1980gc}, we anticipate the fermion condensate to spontaneously break the $U(1)^\prime$ gauge symmetry but preserve the $SU(5)_{\rm GUT}$ symmetry. Defining $Q_{L} = (\psi_{1, L}, \psi_{2,L})$ and similarly for the right-handed fermions, one has 
\beqa
\langle \overline{Q}_{L}Q_{R} \rangle  = \frac{\Lambda_b^3}{16\pi^2} \,\mathbb{I}_{10} \approx 4\pi\, f_\Pi^3 \,\mathbb{I}_{10}  \,, 
\eeqa
which spontaneously breaks the $U(1)^\prime$ gauge symmetry. After $\overline{Q}_{L}Q_{R}$ develops a VEV, the existence of a tadpole potential term for $\varphi$ also generates a nonzero VEV defined as $\langle \varphi \rangle \equiv v_\varphi/\sqrt{2}$, which also breaks $U(1)^\prime$. Specifically, we can write the bare fermion mass matrix $\overline{Q}_L M_Q(\varphi) Q_R$ as
\beqa
M_Q (\varphi) = \left( \begin{array}{cccc}
y_{\rm T} \,\varphi \, \mathbb{I}_3 & 0 & 0 & 0 \\
0 & y_{\rm D} \,\varphi \, \mathbb{I}_2 & 0 & 0 \\
0 & 0 &  y_{\rm T} \, \varphi^{*}\, \mathbb{I}_3 & 0 \\
0 & 0 & 0 & y_{\rm D} \, \varphi^{*}\, \mathbb{I}_2 \\
\end{array} \right) \,.
\eeqa
Using a non-linear parametrization for the big-pions, we have
\beqa
 \overline{Q}_{L}Q_{R}  =\langle \overline{Q}_{L}Q_{R} \rangle\, U \, = \, \Lambda_b \, f_\Pi^2\, \exp\left(\frac{2 i \,T^{A}\Pi^{A}}{f_{\Pi}}\right) \,, 
\eeqa
with the big-pion generators normalized such that $\mbox{Tr}[T^A T^B] = \frac{1}{2} \delta^{AB}$. Thus, the full potential for the scalar field $\varphi$ and the big-pions is
\beqa
V(\varphi, \Pi) = \overline{m}_{\varphi}^{2} \,\varphi^{*} \varphi  + \lambda_{\varphi} (\varphi^{*} \varphi)^{2} \, - \,  \Lambda_b \, f_\Pi^2\, {\rm Tr} \left[ M_{Q}\, U + U^{\dagger} \, M_{Q}^{\dagger} \right] \, .
\label{eq:coupled-potential}
\eeqa
In the limit of $f_\Pi \ll m_{\varphi}$, the  VEV for $\varphi$ induced by the fermion condensation is
\beqa
\langle  \varphi \rangle \equiv \frac{v_{\varphi}}{\sqrt{2}} = \frac{2 \, \Lambda_b\, f_{\Pi}^{2}}{ \overline{m}_{\varphi}^{2}} \, (3\, y_{\rm T} + 2\, y_{\rm D}) \left[1  \, + \, {\cal O}\left(\lambda_{\varphi} (3\, y_{\rm T} + 2\, y_{\rm D})^2 \frac{ \Lambda_b^2\,f_\Pi^4}{ \overline{m}_{\varphi}^{6} } \right) \right]\,.
\eeqa
After fields develop their VEV's, the spontaneous global symmetry breaking pattern is
\beqa
SU(10)_L \times SU(10)_R \times U(1)_V \times U(1)_\varphi  \rightarrow 
SU(10)_V \times U(1)_V \,,
\eeqa
where we have ignored the $U(1)_A$ symmetry that is broken by the $SU(N_b)$ instanton effects. Altogether we anticipate a total of 100 pseudo Nambu-Goldstone bosons (PNGBs), the big-pions. 

In addition to these continuous global symmetries, our model contains two approximate discrete symmetries. Before turning on SM and $U(1)^\prime$ gauge interactions, one can identify the following two transformations:
\beqa
P_m: &&\hspace{-5mm}  \psi^{\rm T, D}_{1} \leftrightarrow \psi^{\rm T, D}_{2}\,, \qquad \varphi \rightarrow  \varphi^* \,, \qquad Z^\prime_\mu \rightarrow -Z^\prime_\mu \,, \qquad T^A A^A_\mu \rightarrow  T^A (A^A_\mu)^{\cal C}  = - (T^A)^* A^A_\mu\,, 
 \\
G_d: &&\hspace{-5mm}\psi^{\rm T, D}_1 \rightarrow (\psi^{\rm T, D}_{2})^{\cal C}  = i \gamma^2 (\psi^{\rm T, D}_2)^* \,, \quad \psi^{\rm T, D}_2 \rightarrow (\psi^{\rm T, D}_{1})^{\cal C}  = i \gamma^2 (\psi^{\rm T, D}_1)^*\,, \quad \varphi \rightarrow  \varphi^* \,, \quad Z^\prime_\mu \rightarrow -Z^\prime_\mu \,.
\eeqa
Here, $\cal C$ denotes charge conjugation and the SM gauge fields are denoted by $A_\mu^A$. All SM fermions are invariant under both discrete transformations. The first symmetry, $P_m$, is just a simple matter parity and is a good symmetry when $y^{\rm T, D}_1 = y^{\rm T, D}_2$. Under it, the SM gauge fields transform as by charge conjugation. The SM fermion electroweak gauge interactions explicitly break $P_m$. The second discrete symmetry, $G_d$, is a new $G$-parity for the new strong dynamics sector~\cite{Bai:2010qg, Antipin:2015xia, Bai:2015nbs, Redi:2016kip}. Because $\psi^{\rm T, D}_{1, L(R)}$ has a different absolute $U(1)^\prime$ charge from $\psi^{\rm T, D}_{2, R(L)}$, the $U(1)^\prime$ gauge interaction explicitly breaks this discrete symmetry.

Decomposing the 100 PNGB's into $[SU(5)_{\rm GUT}\times{U(1)^\prime}]^{P_m\, G_d}$ representations, we have
\beqa
(10 \times 10 - 1) + 1&=& 24_{0}^{++}  + 24_{0}^{--} + 10_{q_2 + q_1} + \overline{10}_{-q_2-q_1}
+ 15_{q_2 + q_1} + \overline{15}_{-q_2-q_1} + 1_{0}^{--} + 1_{0}^{--} \,, 
\eeqa
where $10(\overline{10})$ and $15(\overline{15})$ are not eigenstates of the discrete symmetries. Decomposing  $SU(5)_{\rm GUT}\times{U(1)^\prime}$ into representations of SM gauge groups $[SU(3)_c, SU(2)_W]_{U(1)_Y, U(1)^\prime}$, we have $24_{0}  = (8,1)_{0,0} + (3, 2)_{-5/6, 0} + (\overline{3}, 2)_{5/6, 0} + (1, 3)_{0,0} + (1, 1)_{0, 0}$, $10_{(q_2 + q_1)} = (1, 1)_{1, q_2+q_1} + (\overline{3}, 1)_{-2/3, q_2 + q_1} + (3, 2)_{1/6, q_2 + q_1}$ and $15_{(q_2 + q_1)} = (1, 3)_{1, q_2+q_1} + (3, 2)_{1/6, q_2 + q_1} + (6, 1)_{-2/3, q_2 + q_1}$. The big-pions charged under the SM and $U(1)^\prime$ gauge groups become massive after gauge quantum corrections. We present a calculation of their masses later. Of the remaining four gauge singlets, $(1, 1)_{0, 0}$, three of them are odd under both $P_m$ and $G_d$, while the remaining one is even under both parities. We label the parity-even singlet as $\Pi_{1_A}$, whose generator is 
\beqa
(1, 1)_{0, 0}^{++}: \,\quad \Pi_{1_A} \qquad \mbox{with}  \quad T^{1_A} = \frac{1}{2\sqrt{30} }\, \mbox{diag}( 2\, \mathbb{I}_3, - 3\, \mathbb{I}_2, 2\, \mathbb{I}_3, - 3\, \mathbb{I}_2)\,.
\eeqa
This parity-even singlet will be the lightest state in the spectrum that couples through triangle anomalies to the SM gauge bosons. The other three parity-odd states are
\beqa
(1, 1)_{0, 0}^{--}: &&  \Pi_{1_B} \,, \qquad \mbox{with}  \quad T^{1_B} = \frac{1}{2\sqrt{30} }\, \mbox{diag}( 2\, \mathbb{I}_3, - 3\, \mathbb{I}_2, - 2\, \mathbb{I}_3, 3\, \mathbb{I}_2) \,, \nonumber \\
&&  \Pi_{1_C} \,, \qquad \mbox{with}  \quad T^{1_C} = \frac{1}{\sqrt{20} }\, \mbox{diag}(  \mathbb{I}_5, - \mathbb{I}_5) \,, \nonumber \\
&&  \phi_I \,, \qquad \mbox{with}  \quad  \varphi = \frac{1}{\sqrt{2}}\left( v_\varphi + \phi_R + i \phi_I \right)  \,. 
\eeqa
One linear combination of $\Pi_{1_C}$ and $\phi_I$ will be eaten by the $Z^\prime$ and become its longitudinal component. For the remaining three gauge singlet PNGB's, the Yukawa couplings in Eq.~(\ref{eq:Yukawa-coupling}) explicitly break global $U(1)$'s associated with the PNGB's, making all gauge singlet PNGB's massive.

\subsection{The Big-pion Spectrum}
\label{sec:pion-spectrum}
We first calculate the gauge singlet PNGB spectrum. We expand Eq.~(\ref{eq:coupled-potential}) and find that the $(1, 1)_{0,0}^{++}$ singlet $\Pi_{1_A}$ has no mass mixing with other states; its mass is given by
\beqa
m_{\Pi_{1_A} }^2  = \frac{\sqrt{8}\, \Lambda_b \, v_\varphi }{5}\, \left( 3\, y_{\rm D} \,+\, 2 \, y_{\rm T} \right) \approx
\frac{8 \Lambda_{d}^{2} \, f_{\Pi}^{2}}{5 m_{\varphi}^{2}}\, y_{\rm T}^2 \, (3 + 2\, {\cal R}_y) (2\, + 3\, {\cal R}_y )
\,,
\eeqa
with ${\cal R}_y \equiv y_{\rm D}/y_{\rm T}$. For the three $(1, 1)_{0, 0}^{--}$ singlets, we note that the linear combination of $\Pi_{1_C}$ and $\phi_I$ eaten by $Z^\prime$ is given by its mixing term with $Z^\prime$ via
\beqa
\mathcal{L} \supset  g^\prime  \big(q_{1} - q_{2} \big) Z^\prime_{\mu}\,\partial^{\mu} \left( v_{\varphi}\, \phi_{I} + \sqrt{5} \, f_{\Pi} \, \Pi_{1_C} \right) 
\equiv  g^\prime  \big(q_{1} - q_{2} \big)\, v_{Z^\prime}\, Z^\prime_{\mu}\,\partial^{\mu} \left( \sin{\theta_{Z^\prime}}\, \phi_{I} \, + \, \cos{\theta_{Z^\prime}}\, \, \Pi_{1_C} \right) 
 \, . 
\eeqa
with $v_{Z^\prime}^2 = v_{\varphi}^{2} + 5\, f_{\Pi}^{2}$ and $\sin{\theta_{Z^\prime}}\equiv v_\varphi/ v_{Z^\prime}$. The $Z^\prime$ gauge boson mass is related to the combined VEV from fermion condensation and the VEV of $\varphi$, and is
\beqa
m_{Z'} = g^\prime\,|q_{L}  - q_{R}|\, v_{Z^\prime}\, =\,  g^\prime  \, |q_{L}  - q_{R}| \sqrt{5\, f_{\Pi}^{2} + v_{\varphi}^{2}} \,.
\label{eq:Zprime-mass}
\eeqa
Defining the orthogonal combination to the one eaten by the $Z^\prime$ as $\phi_{I}^\prime =  \cos{\theta_{Z^\prime}}\, \phi_{I} \, - \, \sin{\theta_{Z^\prime}}\, \Pi_{1_C}$, we find the square of the mass mixing matrix in the basis of $(\Pi_{1_B},    \phi_{I}^\prime)^T$ to be
\beqa
\renewcommand{\arraystretch}{1.9}
\left(\Pi_{1_{B}}, \;\phi_{I}^\prime \right)
\left[ \begin{array}{cc}
\frac{2\sqrt{2}\,(3\,y_{\rm D} + 2 \, y_{\rm T} ) }{5} \, v_\varphi\, \Lambda_b & 
\frac{4\sqrt{3}\, (y_{\rm D} \,- \, y_{\rm T} ) }{5} \, v_{Z^\prime} \, \Lambda_b  \\
\frac{4\sqrt{3}\, (y_{\rm D} \,- \, y_{\rm T} ) }{5} \, v_{Z^\prime} \, \Lambda_b &
m_\varphi^2 \left( 1 \,+\, \frac{v_\varphi^2}{5 \, f_\Pi^2 } \right) \\
\end{array} \right]  \, 
\left( \begin{array}{c}
\Pi_{1_{B}} \\
\phi_{I}^\prime \\
\end{array} \right)
\,. 
\eeqa
The rotation matrix from the flavor basis to the mass-eigenstate basis, defined as $\Pi_{1_\beta}$ and $\tilde{\phi}^\prime_I$, is calculated to be
\beqa
\left( \begin{array}{c}
\Pi_{1_{B}} \\
\phi_{I}^\prime \\
\end{array} \right)
= \left( \begin{array}{cc}
\cos{\eta} & -\sin{\eta} \\
\sin{\eta} & \cos{\eta} \\
\end{array} \right) \,
\left( \begin{array}{c}
\Pi_{1_{\beta}} \\
\tilde{\phi}_I \\
\end{array} \right) \, ,
\eeqa
with $\eta \approx 4 \sqrt{\frac{3}{5} } \, \Lambda_{d} \, f_{\Pi}\,(y_{\rm T}-y_{\rm D})/m_{\varphi}^{2} $. The corresponding state masses are
\beqa
m^2_{\Pi_{1_{\beta}} } = 40\,y_{\rm T}\, y_{\rm D}  \, \frac{ \Lambda_{d}^{2} \, f_{\Pi}^{2}}{m_{\varphi}^{2}} \,, \qquad\qquad 
m^2_{\tilde{\phi}_I } = m_{\varphi}^{2} + 8\,(3\, y_{\rm T}^{2} + 2\, y_{\rm D}^{2})\, \frac{\Lambda_{d}^{2}\, f_{\Pi}^{2}}{m_{\varphi}^{2}} \,.
\eeqa

For the gauge charged big-pions, we can use the electromagnetic correction to $\pi^\pm$ in the SM to estimate the radiative corrections from gauge interactions to the big-pion masses. Specifically, we use~\cite{Hill:2002ap}
\beqa
\Delta m^2 = \sum_{i=1,2,3,Z^\prime} \frac{C_2(r_i) \, \alpha_i(f_\Pi) }{\alpha(f_\pi)} \, \frac{f_\Pi^2\, \Delta m_\pi^2}{f_\pi^2} \,.
\eeqa
with $\Delta m_\pi^2 = m_{\pi^\pm}^2 - m_{\pi^0}^2$ and $f_\pi = 93$~MeV. This formula may not work for non-QCD-like strong dynamics. Additional uncertainties on our spectrum calculation would apply in this case. Here, $C_2(r_i)$ is the quadratic Casimir of the representation $r_i$ under the SM $i$'th and $Z^\prime$ gauge groups. To calculate the bare big-quark mass contribution to big-pion masses, we use the Dashen formula 
\beqa
M_{AB}^{2} = \frac{1}{f_{\Pi}^{2}} \langle \overline{Q} \, \{ T^{A},
\{ T^{B}, M_{Q} \} \} Q\, \rangle \, .
\eeqa
The results of various gauge charged big-pion masses are shown in Table~\ref{tab:charged-pion-mass}. 
\begin{table}[htb!]
\renewcommand{\arraystretch}{2.2}
  \centering
  \begin{tabular}{|c|c|}
    \hline \hline
 Big-pions &  $\frac{m^2_{\rm bare}(\Pi^i)} {m^2_{\Pi_{1_A}}}$ \\ \hline
$(8,1)_{0,0}$, $(\overline{3},1)_{-2/3,q_2 + q_1}$, $(6,1)_{-2/3, q_2 + q_1}$   &  $\frac{5}{2 + 3 \, {\cal R}_y }$      \\ \hline
$(1,3)_{0,0}$, $(1,3)_{1,q_2+q_1}$,$(1,1)_{1,q_2+q_1}$   & $\frac{5\,{\cal R}_y }{2 + 3 \, {\cal R}_y }$ \\ \hline
$(3,2)_{-5/6,0}$, $(3,2)_{1/6,q_2+q_1}$    & $\frac{5 ( 1 + {\cal R}_y) }{4 + 6 \, {\cal R}_y}$  \\ 
    \hline \hline
  \end{tabular}
  \caption{The bare big-quark mass contributions to various gauge charged big-pion masses.}  \label{tab:charged-pion-mass}
\end{table}
\begin{figure}[htb!]
\centering
\includegraphics[height=200pt]{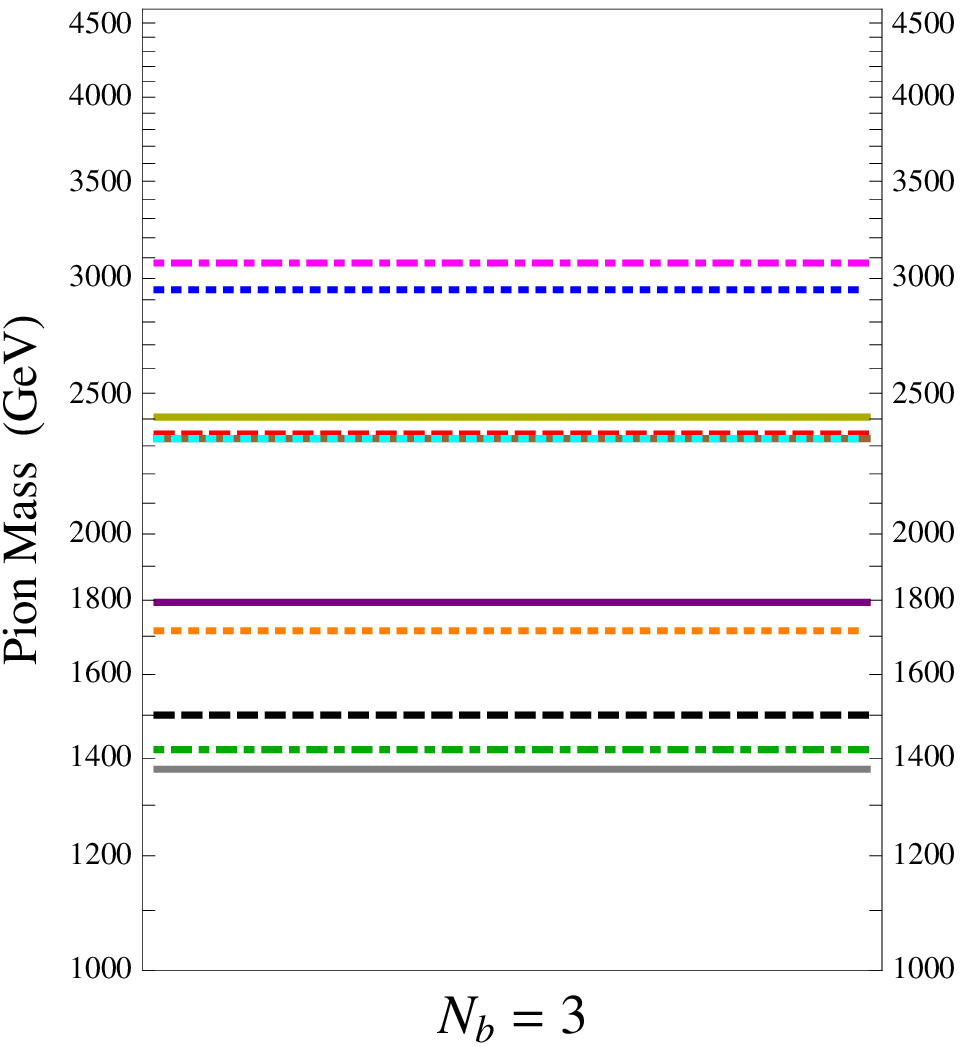}
\hspace{4mm}
\includegraphics[height=202pt]{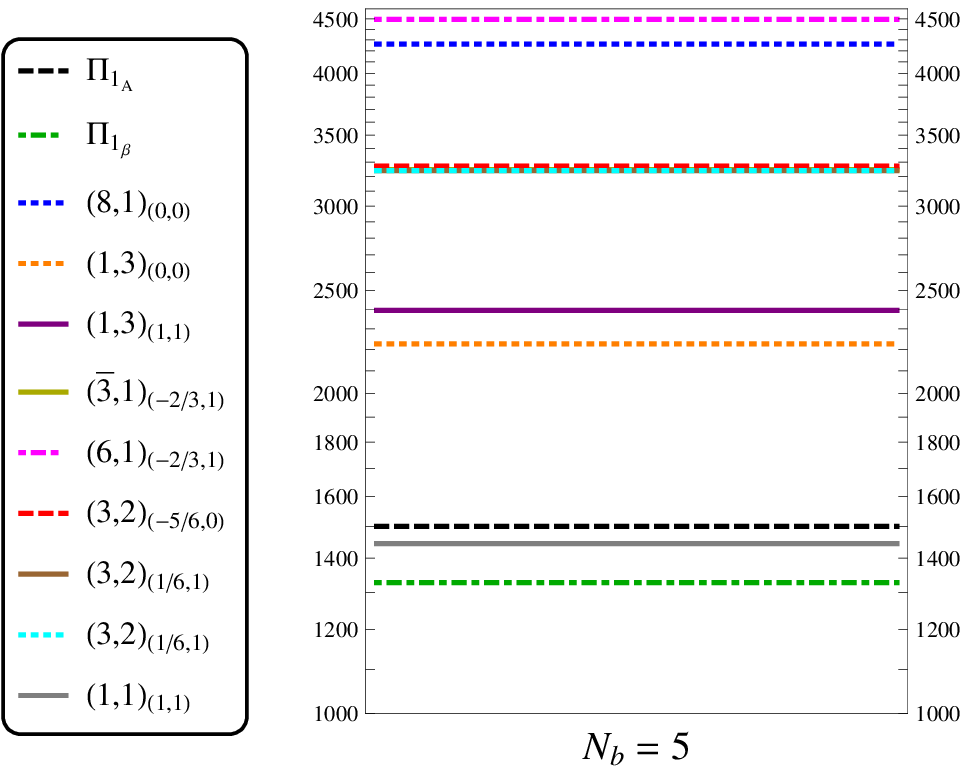}
\caption{Big-pion mass spectrum for $m_{\Pi_{1_A}} = 1500$ GeV and $g^\prime(m_{\Pi_{1_A}})$ = 0.2. The values of $f_\Pi$ used here give a 0.1 fb diphoton cross section for each $N_b$ value.
}
\label{fig:mass_spec}
\end{figure}

Searches for diphoton resonances put the strongest constraints on the model parameters \cite{Khachatryan:2016yec, ATLAS:2016eeo}. To discuss the features of this spectrum, we choose a benchmark mass of $m_{\Pi_{1A}} = 1.5$ TeV for $\Pi_{1_A}$. The ATLAS and CMS collaborations place constraints on the diphoton production cross section for a 1.5 TeV scalar at $\sim 0.2$~fb, so we target a production cross section of $\sigma = 0.1$ fb, which fixes the value of the big-pion decay constant at $f_\Pi = (1040, 1380, 1730)$ GeV for $N_b = (3, 4, 5)$. Then there is only one additional independent combination of Yukawa couplings which we take to be ${\cal R}_y = y_{\rm D} / y_{\rm T}$. As a further benchmark choice, we fix the values of ${\cal R}_y$ for different $N_b$ by choosing an identical boundary condition with $y_{\rm D} = y_{\rm T}$ at the GUT scale. As shown in Appendix~\ref{app:yd-over-yt}, the ratio ${\cal R}_y$ is insensitive to the actual boundary values and ranges from 0.51 for $N_b = 3$ to 0.36 for $N_b = 5$. Fixing the preferred ${\cal R}_y$ values and adding bare big-quark mass and gauge loop contributions to the big-pion masses together, we show the benchmark mass spectra in Fig.~\ref{fig:mass_spec} for $N_b=3$ and $N_b=5$. From Fig.~\ref{fig:mass_spec}, one can see that $\Pi_{1_A}$ is not the lightest big-pion in the spectrum; the gauge singlet big-pion $\Pi_{1_\beta}$ and the big-pion only charged under $U(1)_Y$ and $U(1)^\prime$ are slightly lighter than 1500 GeV. The heaviest big-pion is the color-sextet and has a mass from about 3.1 TeV ($N_b=3$) to 4.5 TeV ($N_b=5$). Following that is the color-octet big-pion with an almost degenerate mass. The radiative corrections from gauge interactions are very important. For instance, the fraction of the QCD-gauge-loop contribution to the color-octet big-pion mass square is around 63\% for $N_b=3$ and 80\% for $N_b=5$. The CP-odd state $\tilde{\phi}_I$ and its CP-even partner $\phi_R$, both of which have much heavier masses, are not shown in Fig.~\ref{fig:mass_spec}. For a fixed value of ${\cal R}_y$ and requiring no Landau pole for the Yukawa coupling $y_{\rm T}$ below the GUT scale, we have an upper bound on the mass parameter $m_\varphi \lesssim  (23, 36, 52)$~TeV
for $N_b = (3, 4, 5)$, after fixing $m_{\Pi_{1A}}$ and $f_{\Pi}$ to our benchmark choices.

\subsection{Interactions and Properties of $Z^\prime$}
\label{sec:zprime-interactions}
As shown in Eq.~(\ref{eq:Zprime-mass}), the $Z^\prime$ mass is related to the chiral symmetry breaking scale of the new strong dynamics sector. For a weak gauge coupling of $g^\prime$ on the order of electromagnetic interaction strength, the $Z^\prime$ mass is anticipated to be $\sim 500~\mbox{GeV}$ for $f_\Pi$ fitting our benchmark parameters. There are no tree-level interactions of the $Z^\prime$ with SM particles. At loop-level, we have found that the kinetic mixing of the $Z^\prime$ with SM hypercharge gauge boson determines its interactions with the SM and decay properties. The relevant kinetic mixing term is defined as 
\beqa
{\cal L} \supset - \frac{1}{4}\hat{B}_{\mu\nu} \hat{B}^{\mu\nu} - \frac{1}{4}\hat{Z}^\prime_{\mu\nu} \hat{Z}^{\prime\,\mu\nu} - \frac{\sin{\chi}}{2}\, \hat{B}_{\mu\nu}\,\hat{Z}^\prime_{\mu\nu} \,,
\eeqa
in the flavor basis. The mixing parameter $\sin{\chi}$ is scale-dependent. Above the GUT scale, $y_{\rm D} = y_{\rm T}$ and $\mbox{Tr}(T_Y \,T_{Z^\prime})=0$, so this coefficient is zero. Below the GUT scale, due to different coupling running of $y_{\rm D}$ and $y_{\rm T}$ (see Appendix~\ref{app:yd-over-yt}), a non-zero value of $\sin{\chi}$ is generated from a vacuum polarization ``bubble digram" with $\hat{B}$ and $\hat{Z}^\prime$ as external fields. At one-loop, the resulting mixing angle is~\cite{Holdom:1985ag}
\beqa
\sin{\chi} = \frac{N_b \, g_Y\, g^\prime}{6\pi^2}(q_1+q_2)\ln{\left( \frac{y_{\rm D} }{y_{\rm T}} \right)} \,.
\eeqa
\begin{figure}[thb!]
\includegraphics[scale=0.4515]{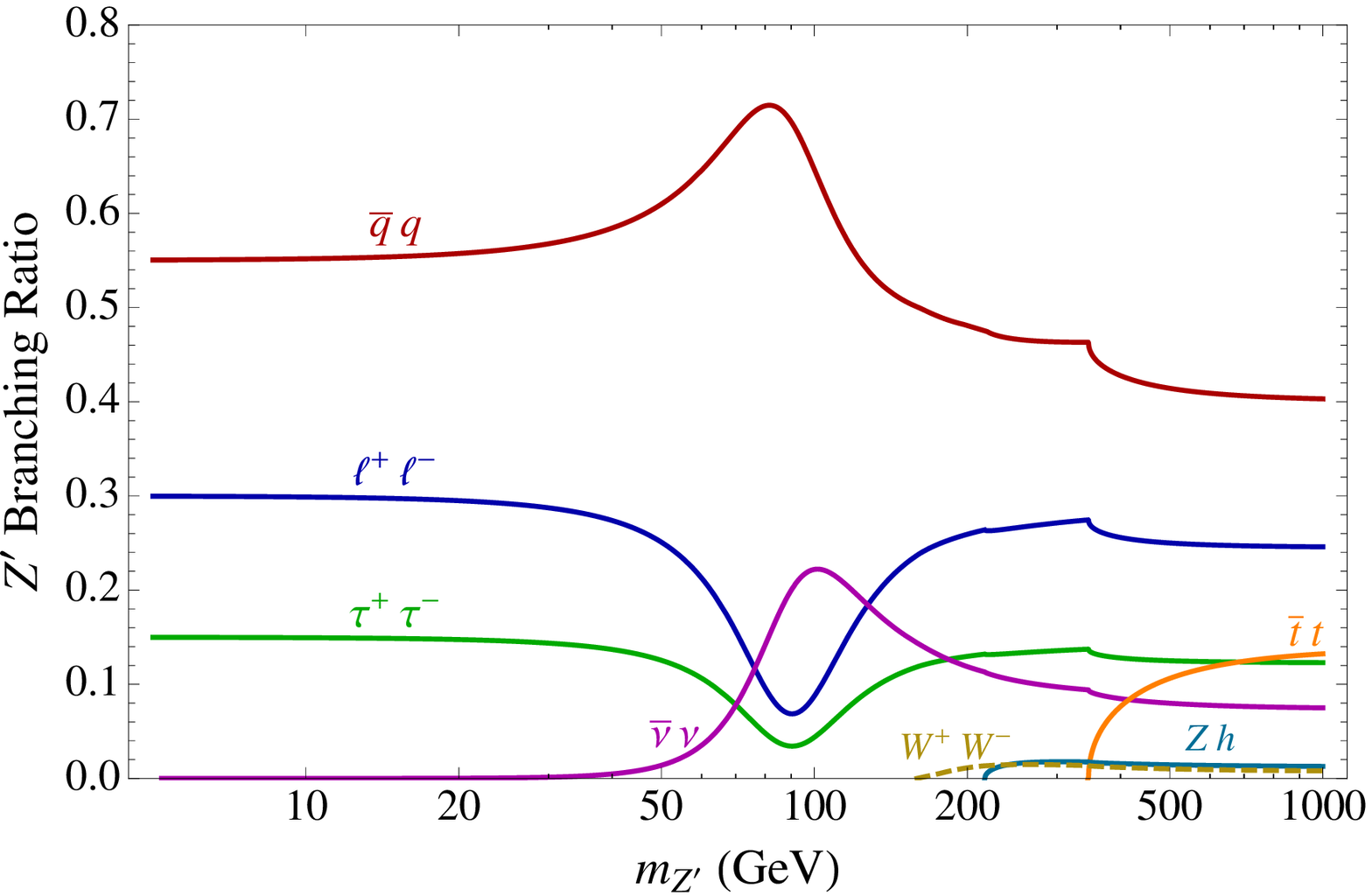} \hspace{3mm}
\includegraphics[scale=0.46]{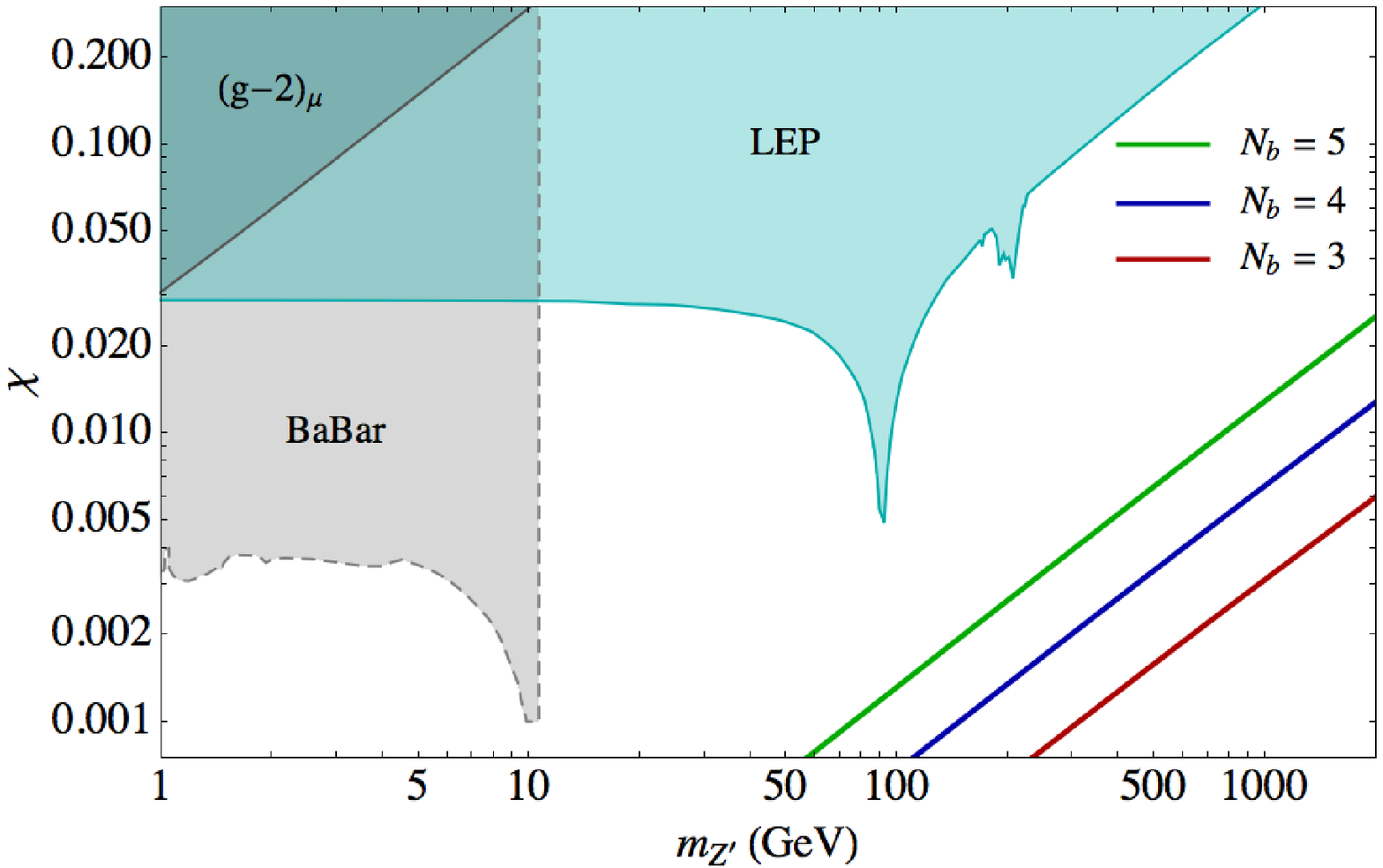}
\centering
\caption{Left panel: the decay branching ratios of $Z^\prime$ to SM particles. Right panel: 95\% CL exclusion limits on $Z^\prime$ kinetic mixing as a function of mass for different $N_b$.
}
\label{fig:zprime-decay}
\end{figure}

In Appendix~\ref{app:Zprime-diagonalize}, we diagonalize both the kinetic and mass mixings of the $Z^\prime$, $\gamma$ and $Z$ boson system. Various decay branching ratios of the $Z^\prime$ are shown in the left panel of Fig.~\ref{fig:zprime-decay}. For the benchmark point of $g^\prime = 0.2$, $N_{d} = 4$ and $f_\Pi =1380$~GeV, the $Z^\prime$ mass is $m_{Z^\prime} = 620$~GeV and its width is $\Gamma_{Z^\prime}\approx 164$~keV. For a lighter $Z^\prime$ below around 50 GeV, the $Z^\prime$ has similar couplings to SM fermions as the SM photon, while for a heavier mass above about 200 GeV, the $Z^\prime$ behaves more like a hypercharge gauge boson. Unless its mass is close to the SM $Z$ boson value, the leptonic branching ratio of this $Z^\prime$ is a factor of $4-5$ larger than the SM $Z$ boson. For the benchmark point of $m_{Z^\prime} = 620$~GeV, the leptonic branching ratio of the $Z'$ is 25\%.

In the right panel of Fig.~\ref{fig:zprime-decay}, we show the bounds on the mass and kinetic mixing of a $Z^\prime$ boson from precision measurements of SM observables~\cite{Hook:2010tw}. Low mass $Z^\prime$ bosons are most strongly constrained by the muon anomalous magnetic moment~\cite{Pospelov:2008zw} and BaBar searches for dark gauge bosons in four-lepton final states~\cite{Aubert:2009af}, whereas high mass $Z^\prime$ bosons are constrained by $e^+ e^-$ collider measurements~\cite{LEP:2003aa}. Not shown are additional narrow-$Z^\prime$ enhanced constraints from various $e^+ e^-$ experiments where $m_{Z^\prime} \simeq \sqrt{s}$ that do not affect our conclusion. We conclude that our model parameter region is unconstrained by low-energy experiments. 

\subsection{Interactions and Properties of Big-Pions}
\label{sec:pion-interaction}
Starting with $\Pi_{1_A}$, we list the most relevant interactions for all big-pions in our model spectrum. Since the triangle anomaly mediated interactions become the leading interactions for the big-pions with real representations under the SM and $U(1)^\prime$ gauge groups, we first introduce a general formula for these types of interactions.  Additional higher-dimensional operators are required to induce decays of the QCD triplet and sextet, as well as $U(1)_Y$ or $U(1)^\prime$ charged big-pions.

For big-pions charged under real representations of SM gauge group, the general form for the triangle anomaly mediated interaction is 
\beqa
\mathcal{L}_{\rm anomaly} \supset -\frac{g_{A}\, g_{B}\, N_b}{64\pi^{2} \, f_{\Pi}} \epsilon^{\mu\nu\rho\sigma}\,\Pi^C F_{\mu\nu}^{A} \, F_{\rho\sigma}^{B} \,d^{ABC}\,, 
\label{eq:anomaly-generaral}
\eeqa
where the group structure constant is $d^{ABC} = 2 \,  {\rm Tr}\left[T^{C}\, \{T^{A},\,T^{B} \}\right]$. Here, the normalization for the big-pion generators is canonical with ${\rm Tr}[T^{C}T^{C^\prime}] = \frac{1}{2}\delta^{C C^\prime}$. The SM gauge group generators are reducible and have $10 \times 10$ representations as $T^{A} = {\rm diag} (t^{A},-t^{A*})$, with ${\rm Tr}(T^{A}T^{B}) = \delta^{AB}$, in the space of the big-pions. 

\subsubsection{Interactions of Parity-even $\Pi_{1_A}$}
\label{sec:pion-interaction-750}
Using the general triangle anomaly interaction formula Eq.~(\ref{eq:anomaly-generaral}), we have interactions for $\Pi_{1_A}$, which is even under both the $P_m$ and $G$ discrete symmetries, shown  in Table~\ref{tab:1A-coupling}. 
\begin{table}[htb!]
\renewcommand{\arraystretch}{2.2}
  \centering
  \begin{tabular}{|c|c c c c c c c|}
    \hline \hline
 Types   &  ${\cal A}_{GG}$  & ${\cal A}_{\gamma \gamma}$ & ${\cal A}_{Z \gamma}$ & ${\cal A}_{Z Z}$ & ${\cal A}_{WW}$ & ${\cal A}_{Z^\prime \gamma}$  & ${\cal A}_{Z^\prime Z}$ \\ \hline
 Couplings     &      $-g_s^2$  & $\frac{7}{3}\,e^2$  & $\frac{ (9\,t_W^{-1} - 5 \, t_W ) }{3}\,e^2$   & $\frac{ (9\,t_W^{-2} + 5 \, t_W^2 ) }{6}\,e^2$ & $\frac{3}{s_W^2}\,e^2$ & $5\,g^\prime e$ &  $-5\,t_W\,g^\prime e$
           \\ 
    \hline \hline
  \end{tabular}
  \caption{The coefficients of $\Pi_{1_A}$ triangle anomaly
    interactions with two gauge bosons as defined by
    $N_b/(16\sqrt{30}\,\pi^2 f_\Pi)\,{\cal A}_{XY}\,
    \Pi_{1_A}\,\epsilon_{\mu\nu\alpha\beta}X^{\mu\nu}
    Y^{\alpha\beta}$. Here, $t_W \equiv \tan{\theta_W}$ and $\theta_W$
    is the Weinberg angle. \label{tab:1A-coupling} }
\end{table}
Based on the interactions of $\Pi_{1_A}$ with gauge bosons, we calculate its various partial widths. For instance, one has 
\beqa
\Gamma(\Pi_{1_A} \rightarrow gg) &=& \frac{8\, N_b^{2}\, g_{s}^{4} \, m_{\Pi_{1_A}}^{3}}{30\times 16^2\,  \pi^{5} \,f_{\Pi}^{2}}   \,, \qquad \qquad
\Gamma(\Pi_{1_A} \rightarrow \gamma\gamma) = \frac{49\, N_b^{2}\, e^{4} \, m_{\Pi_{1_A}}^{3}}{270\times 16^2\, \pi^{5} \, f_{\Pi}^{2}}   \,,   \nonumber \\
\Gamma(\Pi_{1_A} \rightarrow Z^\prime \gamma) &=& \frac{25\, N_b^{2}\, g'^{\, 2} e^{2} \, m_{\Pi_{1_A}}^{3}}{60\times 16^2 \, \pi^{5} \, f_{\Pi}^{2}} \, \left(1-\frac{m_{Z'}^{2}}{m_{\Pi_{1_A}}^{2}}\right)^{3}\,.
\eeqa
There also exist additional interactions of $\Pi_{1_A}$ with other big-pions. For instance, the following charge radius transition operator 
\beqa
c_{Z^\prime}\, \frac{g^\prime}{16\pi^2\, f_\Pi^2} \partial^\mu \Pi_{1_A} \, \partial^\nu \Pi_{1_\beta}\, Z^\prime_{\mu\nu} \,, 
\eeqa
can mediate a sub-dominant decay of $\Pi_{1_A}$ into (off-shell) $\Pi_{1_\beta}$ and $Z^\prime$.

Numerically, we show the various branching ratios and the total width of $\Pi_{1_A}$ in Table~\ref{table:branching-55} for a fixed gauge coupling of $g^\prime = 0.2$ with a corresponding $Z^\prime$ mass of 620~GeV. We can see that the branching ratio into $Z^\prime \gamma$ is comparable to the diphoton one. Depending on the subsequent decays of the $Z^\prime$, one could search for $Z^\prime \gamma$ resonances to confirm this model.
\begin{table}[htb!]
 \renewcommand{\arraystretch}{2.0}
  \centering
\begin{tabular}{|c|c|c|c|c|c|c|c|}
\hline\hline
Mode &  $gg$  & $\gamma\gamma$ &  $Z \gamma$ & $Z Z$ & $W W$ & $Z^\prime \gamma$ & $Z^\prime Z$ \cr \hline
Branching ratio & 0.90 & $0.0046$ & 0.0082 & 0.021 &  0.065 & 0.0025 & 0.0008 \cr \hline
$\Gamma_{\rm tot}$ & \multicolumn{7}{c|}{ $137~\mbox{MeV}\,\left( \frac{N_b}{4} \right)^2 \left( \frac{1380~{\rm GeV} }{f_\Pi} \right)^2$  } \cr
\hline\hline
\end{tabular}
\caption{The branching ratios and total width for $\Pi_{1_A}$ with a mass of 1500 GeV. For decays involving the massive $Z^\prime$ gauge boson, $g' = 0.2$ and $m_{Z^\prime} \approx 620$~GeV for $f_{\Pi} = 1380$~GeV and $N_b = 4$.
}
\label{table:branching-55}
\end{table}

\begin{figure}[htb!]
  \centering
  \includegraphics[scale=0.8]{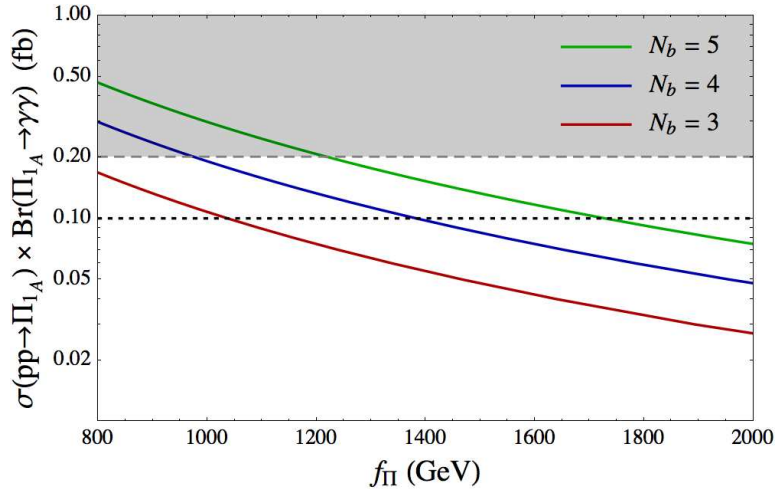}
  \caption{The diphoton rate as a function of $f_\Pi$ for different $N_b$. The shaded region shows the exclusion limit from the ATLAS diphoton search at our benchmark mass of 1.5 TeV~\cite{ATLAS:2016eeo}. 
  \label{fig:diphoton-fpi}}
\end{figure}

Experimental signatures of this model may first appear in the diphoton channel. Using its couplings to two gluons and two photons, we compute the production cross section for $gg \rightarrow \Pi_{1_A} \rightarrow \gamma\gamma$, which in the narrow width approximation is given by 
$\sigma(gg \rightarrow \Pi_{1_A} \rightarrow \gamma\gamma) =  \sigma (gg \rightarrow \Pi_{1_A}) \times {\rm Br}(\Pi_{1_A} \rightarrow \gamma\gamma)$, with
\begin{equation}
 \sigma (gg \rightarrow \Pi_{1_A}) = \frac{N_b^{2}\, g_{s}^{4}\, m_{\Pi_{1_A}}^{4}}{30\times 16^2 \, \pi^{4} \, f_{\Pi}^{2}}\, \frac{\pi}{\hat{s}}\delta(\hat{s}-m_{\Pi_{1_A}}^{2}) \,.
 \label{eq:singlet-product}
\end{equation}
Here, $\hat{s}$ means the center-of-mass energy of partons. Integrating this cross section with the MSTW2008 NNLO central parton distribution function set~\cite{Martin:2009iq} and an NNLO K-factor of 2.5~\cite{Catani:2003zt} for a 1500 GeV big-pion at the 13 TeV LHC, we have the required $f_\Pi$ as
\begin{equation}
f_\Pi = 1380 \, {\rm GeV} \, \left(\frac{N_b}{4}\right) \, \left( \frac{0.1\, {\rm fb}}{\sigma\times{\rm Br}}\right)^{1/2}  \,.
\end{equation}
In Fig.~\ref{fig:diphoton-fpi}, we show the diphoton rate as a function of $f_\Pi$ for different big-colors $N_b$. The decay constants fitting our benchmark diphoton cross section of 0.1 fb range from roughly 1 TeV to 1.7 TeV, so the related confinement scale $\Lambda_b \sim 4\pi f_\Pi$ varies from 13 TeV to 22 TeV. The heavier states including vector $\rho_b$, $a_b$ mesons and baryons are heavy and unlikely to be probed by the LHC Run 2. The baryons in our model either decay via higher dimension operators or are stable and form a component of the dark matter. We don't explore their properties in detail in this paper.

In our chiral composite model, $\Pi_{1_A}$ decays to $Z^\prime \gamma$ with a similar branching fraction to $\gamma \gamma$. For our benchmark point with $g^\prime = 0.2$ and $m_{Z^\prime} = 620$~GeV, the production cross section times branching ratio of the final state of $\ell^+ \ell^- \gamma$ is
\beqa
\sigma (pp \rightarrow \Pi_{1_A}) \, \times\,\mbox{Br}( \Pi_{1_A} \rightarrow Z^\prime \gamma)\,\times\, \mbox{Br}(Z^\prime \rightarrow \ell^+ \ell^-)  \approx \left(\frac{g^\prime}{0.2}\right)^2\times 0.013~\mbox{fb}\,,
\eeqa
at the 13 TeV LHC. While this channel is unlikely to provide the first hints of new physics from this model, observation of such a decay mode serves as a key indicator of new chiral dynamics in the big-color sector.

\subsubsection{Interactions of Parity-odd $\Pi_{1_\beta}$}
\label{sec:pion-interaction-1_beta}
A single parity-odd big-pion, $\Pi_{1_\beta}$, cannot couple to two gauge bosons because of the discrete symmetries. From a box diagram at one loop, we expect it to couple to three gauge bosons. For instance, one can have the following dimension-9 operator
\beqa
&&c^G_{\beta} \, \frac{g_s^2 \, g^\prime}{16\pi^2\, f_\Pi\,\Lambda_b^4} \, \partial_\mu \Pi_{1_\beta}\, G^a_{\mu\rho} \, \widetilde{G}^{a\,\rho}_{\sigma}\, \partial_\nu Z^{\prime\,\nu\sigma}
\,,
\eeqa
and similar interactions with the $W^i$ and $B$ gauge bosons. So, the leading decay channel of $\Pi_{1_\beta}$ is $\Pi_{1_\beta} \rightarrow g g Z^\prime$. The dimension-7 operator like $\Pi_{1_\beta} G^a_{\mu\rho}\widetilde{G}^{a\, \rho}_\sigma Z^\prime_{\mu \sigma}$ can be shown to be zero.  There also exist operators containing two $\Pi_{1_\beta}$'s, for instance
\beqa
c^{GG}_\beta\, \frac{g_s^2}{16\pi^2\,f_\Pi^2} \, \Pi_{1_\beta}\,\Pi_{1_\beta}\, G^a_{\mu\nu} \, G^{a\, \mu\nu} \,,
\eeqa
with $c^{GG}_\beta$ coming from the strong dynamics and of order unity. This operator provides the dominant interaction for producing $\Pi_{1_\beta}$ at the LHC. For $c^{GG}_\beta =1$, $f_\Pi = 1040$~GeV, and $m_{\Pi_{1_\beta}} = 1400$~GeV, the production cross section of $p p \rightarrow \Pi_{1_\beta}\Pi_{1_\beta}$ is 0.005~fb at the 13 TeV LHC, which is unlikely to be observed at the LHC Run 2. After both $\Pi_{1_\beta}$'s decay, we have a very interesting signature with four jets plus one or two leptonic $Z^\prime$.

\subsubsection{Interactions of Parity-even Color-octet $\Pi_8$}
\label{sec:pion-interaction-octet-even}
The parity-even color-octet can also couple to two SM gauge bosons through triangle anomalies. Because of the QCD gauge invariance, it couples to two gluons or one gluon plus one hypercharge boson. After electroweak symmetry breaking, the relevant interactions are
\beqa
\mathcal{L}_{\rm anomaly} & \supset& -\frac{1}{\sqrt{2}}\frac{N_{b}\, g_{s}^{2}}{32\pi^{2} f_{\Pi}} \, d^{abc}\, \Pi_{8}^{a} \, \epsilon^{\mu\nu\rho\sigma} \, G_{\mu\nu}^{b} \, G_{\rho\sigma}^{c} - \frac{2\sqrt{2}}{3} \frac{N_{b}\, g_{s} \, e}{32\pi^{2} f_{\Pi}} \, \Pi_{8}^{a} \, \epsilon^{\mu\nu\rho\sigma} \,G_{\mu\nu}^{a} \, F_{\rho\sigma} 
\nonumber \\
&& +\,\frac{2\sqrt{2}}{3} \frac{N_{b}\, g_{s} \, e\, t_W }{32\pi^{2} f_{\Pi}} \, \Pi_{8}^{a} \, \epsilon^{\mu\nu\rho\sigma} \,G_{\mu\nu}^{a} \, Z_{\rho\sigma} -\sqrt{2}\, \frac{N_{b}\, g_{s} \, g'}{32\pi^{2} f_{\Pi}}\, \Pi_{8}^{a} \, \epsilon^{\mu\nu\rho\sigma} G_{\mu\nu}^{a} \, Z'_{\rho\sigma} \, .
\eeqa
After summing color factors, the partial widths of $\Pi_8$ are given by
\beqa
\Gamma(\Pi_{8} \rightarrow gg) &=& \frac{25}{4} \frac{N_{b}^{2}\, g_{s}^{4} \, m_{\Pi_8}^{3}}{30\times 16^2 \, \pi^{5} \, f_{\Pi}^{2}} \,, \qquad\qquad
\Gamma(\Pi_{8} \rightarrow g\gamma) = \frac{8\,\alpha}{15\,\alpha_s}\, \Gamma(\Pi_{8} \rightarrow gg) \,, \nonumber \\
\Gamma(\Pi_{8} \rightarrow gZ) &= & \frac{8 \, \alpha\, t_{W}^2 }{15 \,  \alpha_s }  \, \Gamma(\Pi_{8} \rightarrow gg) \,,\qquad\qquad 
\Gamma(\Pi_{8} \rightarrow gZ^\prime) = \frac{6\, g'^{2}}{5\, g_{s}^{2}}  \, \Gamma(\Pi_{8} \rightarrow gg) \,.
\eeqa
Numerically, we show the various decay branching ratios and the total width in Table~\ref{table:branching-octet}.  
\begin{table}[h!]
 \renewcommand{\arraystretch}{2.0}
  \centering
\begin{tabular}{|c|c|c|c|c|}
\hline\hline
Mode &  $gg$  & $gZ'$ &  $g \gamma$ & $g Z$ \cr \hline
Branching ratio & 0.91 & $0.038$ & 0.042 & 0.013 \cr \hline
$\Gamma_{\rm tot}$ & \multicolumn{4}{c|}{ $1.45~\mbox{GeV}\,\left( \frac{N_d}{4} \right)^2 \left( \frac{1380~\mbox{GeV} }{f_\Pi} \right)^2$ }  \cr
\hline\hline
\end{tabular}
\caption{The branching ratios and total width for $ \Pi_{8}$ with mass 3580 GeV. For decays involving the massive $Z'$ gauge boson, $g' = 0.2$ was used. All phase space and polarization factors have been neglected here due to the large $ \Pi_{8}$ mass.}
\label{table:branching-octet}
\end{table}

At the LHC, the color-octet big-pion can be singly produced from two gluons. The parton-level production cross section is
\begin{equation}
 \sigma (gg \rightarrow \Pi_{8}) = \frac{25}{4}\frac{N_b^{2}\, g_{s}^{4}\, m_{\Pi_8}^{4}}{30\times 16^2 \, \pi^{4} \, f_{\Pi}^{2}}\, \frac{\pi}{\hat{s}}\delta(\hat{s}-m_{\Pi_8}^{2}) \, .
\end{equation}
It is also interesting to compare the above formula to the color-singlet production in Eq.~(\ref{eq:singlet-product}). The ratio of the two cross sections is
\beqa
\frac{\sigma (gg \rightarrow \Pi_{8})}{  \sigma (gg \rightarrow \Pi_{1_A})} = \frac{25}{4} \, \frac{m_{\Pi_8}^{2}\,\delta(\hat{s}-m_{\Pi_8}^{2}) }{m_{\Pi_{1_A}}^{2}\,\delta(\hat{s}-m_{\Pi_{1_A}}^{2})} \,,
\eeqa
which is independent of $f_\Pi$ and $N_d$. 

\begin{figure}[htb!]
\includegraphics[scale=0.45]{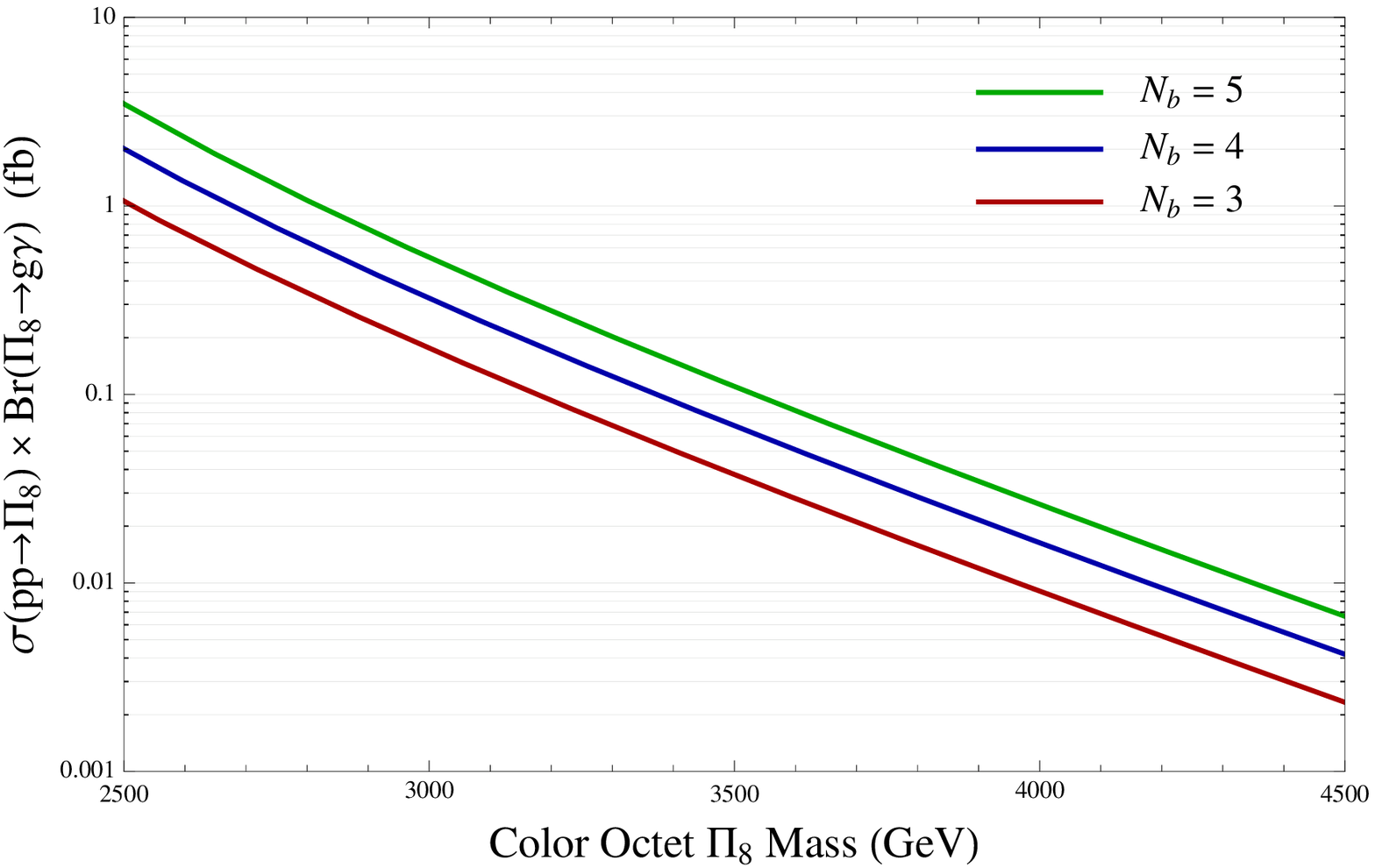}
\includegraphics[scale=0.45]{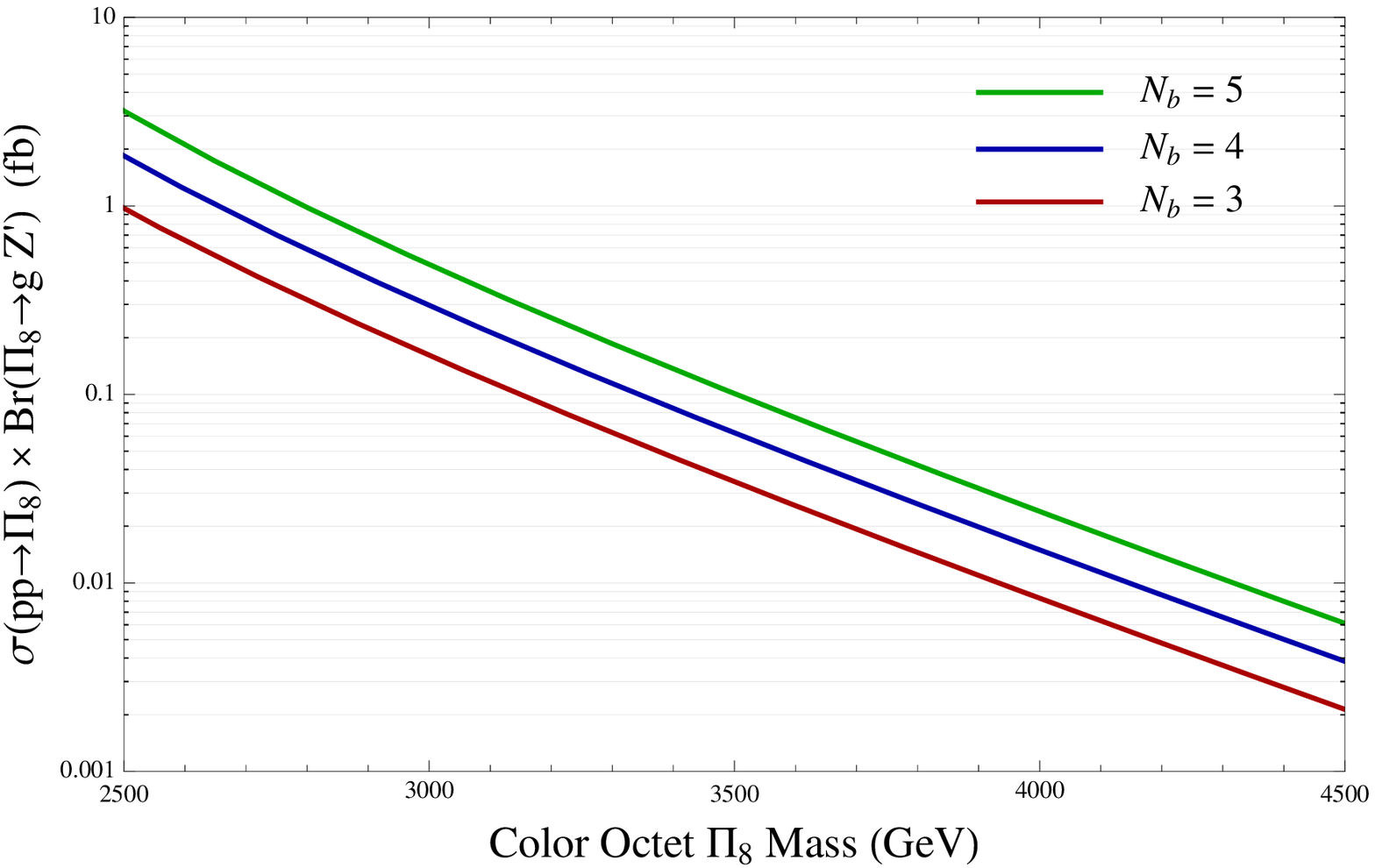}
\centering
\caption{Color-octet big-pion $\Pi_{8} \equiv (8,1)_{0,0}$ single production cross-section times its branching ratio to $g\gamma$ (left panel) and $gZ'$ (right panel) as a function of mass at the 13 TeV LHC. This color-octet big-pion is even under both parities, namely $(P_{m}, G_{d}) = (+,+)$. An NNLO K-factor of 3.0 was used~\cite{Idilbi:2009cc}. 
}
\label{fig:m8_XsecBr}
\end{figure}

Integrating this parton-level cross section with the MSTW2008 NNLO central parton distribution function set and an NNLO K-factor of 3.0~\cite{Idilbi:2009cc} for the benchmark mass 3580 GeV ($f_{\Pi} \sim 1380$ GeV and $N_b = 4$) color-octet at the 13 TeV LHC, we find $\sigma(gg \rightarrow \Pi_{8} \rightarrow gZ') \sim 0.05 \, {\rm fb}$ for $g'=0.2$. For different masses, we show the color-octet single production cross section times branching ratio in Fig.~\ref{fig:m8_XsecBr} for $g\gamma$ (left panel) and $gZ^\prime$ (right panel). As already studied in Ref.~\cite{Bai:2016czm}, the dijet and jet plus photon resonance searches have already imposed stringent constraints on the color-octet production cross sections. However, color-octet pions from the benchmark points considered here are unconstrained by such searches.

Different from the pure vector-like models in Ref.~\cite{Bai:2016czm}, one can also search for the color-octet big-pion as a three-body resonance of $j\,\ell^+ \ell^-$ with a mass between 3 TeV to 4.3 TeV. The signal cross section times branching ratio is 0.013 fb for a 3580 GeV color-octet pion which, while unlikely to be observed in the near future, provides a relatively clean signal for high luminosity LHC. The color-octet big-pions can also be pair produced via their QCD interactions (see Refs.~\cite{Bai:2010dj,Chivukula:2016zbe,Chivukula:2013xka} for similar phenomenology studies). The leading discovery channel is a pair of dijet resonances, but the semi-weak decays of one octet big-pion into $g\gamma$ and $g Z^\prime$ should be visible as well. 

\subsubsection{Interactions of Parity-odd Color-octet $\Pi_8^{\rm odd}$}
\label{sec:pion-interaction-octet-odd}
The discrete-symmetry-odd color-octet big-pion $\Pi_8^{\rm odd}$ can decay into three gauge bosons via the following dimension-7 operator 
\beqa
c^{{\rm o}\,GG}_8\, \frac{g_s^2\,g^\prime}{16\pi^2\,f_\Pi\,\Lambda_b^2} \, f^{abc}\, \Pi^{{\rm odd}\, a}_8 \, G^b_{\mu\rho} \,\widetilde{G}^{c\,\rho}_{\nu}\,Z^{\prime\, \mu\nu} \,,
\qquad
c^{{\rm o}\,GB}_8\, \frac{g_s\,g_Y\,g^\prime}{16\pi^2\,f_\Pi\,\Lambda_b^2} \,\Pi^{{\rm odd}\, a}_8 \, G^a_{\mu\rho}\, \widetilde{B}^{c\,\rho}_{\nu}\,Z^{\prime\, \mu\nu} 
 \,.
\eeqa
Furthermore, it can also decay into the lighter discrete-symmetry-odd singlet $\Pi_{1_\beta}$ via the dimension-6 operators
\beqa
c^{{\rm o}\,GG}_{8\beta}\, \frac{g_s^2}{16\pi^2\,f_\Pi^2} \, d^{abc}\, \Pi^{{\rm odd}\, a}_8\,\Pi_{1_\beta} \, G^b_{\mu\nu}\, G^{c\,\mu\nu} \,,\qquad
c^{{\rm o}\,GB}_{8\beta}\, \frac{g_s\,g_Y}{16\pi^2\,f_\Pi^2} \,  \Pi^{{\rm odd}\, a}_8\,\Pi_{1_\beta} \, G^a_{\mu\nu}\, B^{\mu\nu}  \,,
\eeqa
which should provide the leading decaying channels $\Pi^{\rm odd}_8 \rightarrow \Pi_{1_\beta} + 2 g/g\gamma/gZ$. Similarly, $\Pi^{\rm odd}_8$ can also decay into $\Pi_{1_A}$ via
\beqa
c^{{\rm o}\,GZ'}_{8A}\, \frac{g_s\,g'}{16\pi^2\,f_\Pi^2} \,  \Pi^{{\rm odd}\, a}_8\,\Pi_{1_A} \, G^a_{\mu\nu}\, Z^{\prime\,\mu\nu}  \,,
\eeqa
which provides the subdominant decay channel $\Pi^{\rm odd}_8 \rightarrow \Pi_{1_A} \, g \, Z'$. 

At the 13 TeV LHC, $\Pi_8^{\rm odd}$ could be pair-produced via its QCD interactions. The tree-level production cross section is around $2.3\times 10^{-4}$($7.8\times 10^{-6}$) fb for a 3.0(3.6) TeV $\Pi_8^{\rm odd}$, which is too small to be observed at the LHC.

\subsubsection{Interactions of Weak Triplets $\Pi_3$ and $\Pi_3^{\rm odd}$}
\label{sec:pion-interaction-weak-triplet}
The discrete-symmetry-even weak-triplet $\Pi_3$ contains both electric charged big-pions $\Pi_3^\pm$ and a neutral big-pion $\Pi_3^0$. Through the triangle anomaly, they can couple to two gauge bosons via
\beqa
\mathcal{L}_{\rm anomaly} &\supset& \sqrt{2} \frac{N_{b}\, e}{32\pi^{2}\,s_W\, f_{\Pi}} \, \epsilon^{\mu\nu\rho\sigma}\, (\Pi_{3}^{+} \,W_{\mu\nu}^{-} + \Pi_{3}^{-}\,W_{\mu\nu}^{+}) (e\,F_{\rho\sigma} - e\,t_W\,Z_{\rho\sigma} - g'\,Z'_{\rho\sigma})   
\nonumber \\
&& +\sqrt{2} \frac{N_{b}\, e^{2}}{32\pi^{2} f_{\Pi}} \, \epsilon^{\mu\nu\rho\sigma}  \,\Pi_{3}^{0} \, \left[ (t_W^{-1}-t_W)\,Z_{\mu\nu}\,F_{\rho\sigma} + F_{\mu\nu}\,F_{\rho\sigma} - Z_{\mu\nu}\,Z_{\rho\sigma}\right] 
\nonumber \\
&& -\sqrt{2}\, \frac{N_{b}\, g'\, e}{32\pi^{2} f_{\Pi}}\,\epsilon^{\mu\nu\rho\sigma}  \,\Pi_{3}^{0} \,(F_{\mu\nu} \,Z'_{\rho\sigma}+ t_W^{-1} Z_{\mu\nu}\, Z'_{\rho\sigma}) \,.
\eeqa
We show numerical values for the various branching ratios in Table~\ref{table:branching-weak-trip} for the benchmark model point with a mass of 970~GeV. At the LHC, the weak-triplet big-pions can be singly produced from vector-boson fusion with two forward jets and small cross sections. They can also be produced in pairs from their weak interactions. While the weak triplets are unlikely to be observed at the LHC, a future 100~TeV collider could be capable of probing these states.

\begin{table}[h!]
 \renewcommand{\arraystretch}{2.0}
 \hspace{0.1cm}
\begin{tabular}{|c|c|c|c|c|c|}
\hline\hline
$\Pi_{3}^{0}$ decay &  $\gamma\gamma$  & $ZZ$ &  $Z \gamma$ & $Z' \gamma$ & $Z' Z$ \cr \hline
Br & 0.28 & $0.28$ & 0.204 & 0.057 & 0.179 \cr \hline
$\Gamma_{\rm tot}$ & \multicolumn{5}{c|}{ $14~\mbox{MeV}\,\left( \frac{N_d}{4} \right)^2 \left( \frac{1380~\mbox{GeV} }{f_\Pi} \right)^2$}   \cr
\hline\hline
\end{tabular}
\hspace{1cm}
\begin{tabular}{|c|c|c|c|}
\hline\hline
$\Pi_{3}^{\pm}$ decay &  $W^\pm \gamma$  & $W^\pm Z$ &  $W^\pm Z'$ \cr \hline
Br & 0.579 & $0.186$ & 0.236 \cr \hline
$\Gamma_{\rm tot}$ & \multicolumn{3}{c|}{ $14~\mbox{MeV}\,\left( \frac{N_d}{4} \right)^2 \left( \frac{1380~\mbox{GeV} }{f_\Pi} \right)^2$}  \cr
\hline\hline
\end{tabular}
\caption{The decay branching ratios and total width for $\Pi_{3}^{0}$ (left panel) and $\Pi_{3}^{\pm}$ (right panel) with a mass of 1960 GeV. For decays involving the massive $Z'$ gauge boson, $g' = 0.2$ was used.}
\label{table:branching-weak-trip}
\end{table}

For the discrete-symmetry-odd weak-triplet $\Pi_3^{\rm odd}$ and similar to the color-octet case, the leading decaying operators are
\beqa
c^{{\rm o}\,WB}_{3\beta}\, \frac{e^2}{16\pi^2\,s_W\,c_W\,f_\Pi^2} \, \Pi^{{\rm odd}\, i}_3\,\Pi_{1_\beta} \, W^i_{\mu\nu}\, B^{\mu\nu} \,, \qquad 
c^{{\rm o}\,WZ'}_{3A}\, \frac{e\,g'}{16\pi^2\,s_W\,f_\Pi^2} \, \Pi^{{\rm odd}\, i}_3\,\Pi_{1_A} \, W^i_{\mu\nu}\, Z^{\prime\,\mu\nu}  \,.
\eeqa
So, the main decay channels for the charged states are $\Pi^{{\rm odd}\,\pm}_3\rightarrow \Pi_{1_\beta} W^\pm Z/\gamma$ and  $\Pi_{1_A} W^\pm Z^\prime$.

\subsubsection{Additional UV Interactions of Complex Big-Pions}
\label{sec:UV-complex}
Additional interactions in the UV physics are required to make the big-pions with complex representations under SM gauge groups decay~\cite{Buttazzo:2016kid}. In general, there are two classes  of operators, depending on whether additional $\varphi$ insertions are needed or not. If we write down such operators in a GUT-preserving form, then the gauge structure of the operator is fixed. The big-pions with complex SM representations can come from either the decomposition of $5 \times 5$ (and $ \overline{5} \times \overline{5}$ which we neglect as it can be treated analogously by conjugation) or $5 \times \overline{5}$. To allow all complex big-pions to decay, we need to introduce operators in which the SM fields transform as $ \overline{10}$, $ \overline{15}$ and $24$. The $ \overline{10}$ can come from $\overline{5} \times  \overline{5}$ or $5 \times 10$. Both cases lead to one big-pion that decays as a leptoquark and one big-pion that decays as a diquark. The $ \overline{15}$ must come from a product of $ \overline{5} \times \overline{5}$ SM fermions, such that the $(6,1)_{-2/3,q_1+q_2}$ decays as down-type diquarks, while the $(\bar{3},2)_{1/6,q_1+q_2}$ behaves as a leptoquark. The QCD-neutral complex big-pions decay to leptons. The operator with SM fields in a $24$ can come from a $5 \times \overline{5}$ or a $10 \times \overline{10}$. In either case, it decays as a diquark. For complex big-pions that also have $U(1)^\prime$ charge, additional insertions of the $\varphi$ field are required. If we choose $q_1 = 1$ and $q_2 = 0$, then only one insertion is required and the UV operators inducing the decays are dimension-7. In order to have $(3, 2)_{-5/6,0}$ decay, one also needs to add a dimension-6 operator without $\varphi$ insertion. For example, one could introduce decays for all complex big-pions with the following operators,
\beqa
\frac{\varphi^* \overline{\psi}_{2 i} \gamma^5 \psi_{1 j}  \overline{5}_i P_L  \overline{5}_j}{\Lambda^{3}_{1}}\,, \qquad 
\qquad \frac{\overline{\psi}_2 \gamma^\mu \gamma^5 T^A \psi_2\, \overline{5} \gamma^\mu P_R T^A 5}{\Lambda^{2}_{2}}\, ,
\eeqa
where $i,j$ are indices of $SU(5)$ and we suppress Lorentz and flavor indices. The new particles required to UV-complete the above two operators may change gauge coupling running if they are not $SU(5)$ singlets.

These operators break the discrete symmetries of the theory and so induce decays for both even and odd big-pions. To have these big-pions decay before Big Bang Nucleosynthesis ($\sim1$~s), we estimate that the cutoff scales should be less than $\mathcal{O}(10^{7} {\rm \, GeV})$ and $\mathcal{O}(10^{10} {\rm \, GeV})$, for $\Lambda_{1}$ and $\Lambda_{2}$ respectively. If the abundance of the complex big-pions is small, their lifetime could be longer and leads to weaker constraints on cutoffs~\cite{Harigaya:2016pnu}. Searches for stopped long-lived particles at the 8 TeV LHC place a bound on the color triplet mass of to be above around 470~GeV, for a wide range of decay times, $10^{-6}\, {\rm s} \lesssim \tau \lesssim 10^{4} \, {\rm s}$~\cite{Khachatryan:2015jha}. Similarly for the color sextet, the bound is that the mass should be above around 690~GeV. Furthermore, for $\tau$ above  ${\cal O}(10~\mbox{ns})$, the searches for long-lived charged particles have imposed a more stringent bound, which requires the color triplet complex scalar mass above around 900 GeV~\cite{Chatrchyan:2013oca}. This constraint can be easily satisfied in our model, as can be seen in Fig.~\ref{fig:mass_spec}. Since the colored big-pions in our model are much heavier, we do not anticipate any constraints coming from these bounds. We show the summary table of all big-pion decays in Table~\ref{tab:pion_decay_modes}. 

\begin{table}[htb!]
\centering
{\small
\renewcommand{\arraystretch}{2.0}
  \begin{tabular}{|c|c|}
    \hline \hline
 Big-Pion &  Decay Modes\\ \hline
$\Pi_{1_{A}}$ &  $gg$, $\gamma \gamma$, $Z\gamma$, $ZZ$, $WW$, $Z^\prime\gamma$,  $Z^\prime Z$    \\ \hline
$\Pi_{1_{\beta}}$ &  $ggZ'$      \\ \hline
$(8,1)_{0,0}^{++}$ &  $gg$, $g\gamma$, $g Z$, $g Z'$      \\ \hline
$(8,1)_{0,0}^{--}$ &  $\Pi_{1_{\beta}}(gg, g\gamma, gZ)$, $\Pi_{1_{A}}\,g\,Z'$      \\ \hline
$(1,3)_{0,0}^{++}$  & $W\gamma$, $WZ$, $\gamma\gamma$, $\gamma Z$, $ZZ$, $\gamma Z'$, $Z Z'$, $W Z'$\\ \hline
$(1,3)_{0,0}^{--}$  & $\Pi_{1_{\beta}}(W\gamma,  WZ, \gamma \gamma, ZZ, Z\gamma)$, $\Pi_{1_A}(WZ', \gamma Z', Z Z')$ \\ \hline
    \hline
  \end{tabular}
    \begin{tabular}{|c|c|}
    \hline \hline
 Big-Pion &  Decay Modes\\ \hline
$(6,1)_{-2/3, \, q_2 + q_1}$   &  di-quark      \\ \hline
$(1,3)_{1,\, q_2+q_1}$  & di-lepton \\ \hline
$(1,1)_{1, \, q_2+q_1}$   & di-lepton \\ \hline
$(\overline{3},1)_{-2/3, \, q_2 + q_1}$   &  di-quark, leptoquark      \\ \hline
$(3,2)_{-5/6,\, 0}$    & leptoquark  \\ \hline
$(3,2)_{1/6, \, q_2+q_1}$    & di-quark, leptoquark  \\ 
    \hline \hline
  \end{tabular}
  }
  \caption{A summary of all big-pion decays. Left: big-pions in real representations with collider prompt decay modes. Right: big-pions in complex representations which may live long enough to be collider stable.}\label{tab:pion_decay_modes}
\end{table}
%

\section{Discussion and Conclusions}
\label{sec:conclusion}
In this article we have examined signatures of a chiral composite model in the context of LHC Run 2 discovery. To avoid spoiling the unification of the SM gauge couplings we require that the new matter content be embedded in complete $SU(5)$ GUT representations. The chiral structure is then necessitated by the desire to explain the small masses of the constituent fermions relative to the GUT scale. While recent searches for diphoton resonances by ATLAS and CMS have placed strong constraints on the production cross section of the lightest state coupling to SM gauge bosons, we find that regions of parameter space exist where diphoton signatures would be observable by the end of Run 2. For the benchmark choice of $m_{\Pi_{1_A}} = 1.5$~TeV and $f_\Pi = 1380$~GeV for $N_b = 4$, a production cross section times branching ratio to two photons of $\sigma \times \text{Br} \sim 0.1$~fb could be observable in the near future.

The model also predicts a rich spectrum of pseudo-scalars accompanying the potential digamma resonance. Among the unique features of this model, it predicts two very interesting decay channels involving the $Z^\prime$ gauge boson. One is the decay of $\Pi_{1_A}$ to $Z^\prime \gamma$ and the other is the decay of the color-octet big-pion to $g Z^\prime$. For our benchmark model point with $g^\prime = 0.2$ and $m_{Z^\prime} = 620$~GeV, the $Z^\prime$ has roughly a 25\% branching fraction to a di-lepton final state, giving signatures of $\ell^{+} \ell^{-} \gamma$ or $\ell^{+} \ell^{-} j$. If a scalar resonance is confirmed in the future, these signals could be considered smoking gun signatures for chiral composite models.

\subsection*{Acknowledgments}
We would like to thank Vernon Barger for discussion. This work is supported by the U. S. Department of Energy under the contract DE-FG-02-95ER40896.

\appendix
\section{Light Elementary Scalar Field}
\label{app:real-scalar}
The model presented in the main paper includes a chiral symmetry-breaking scalar whose mass is expected to be out of reach of near-future experiments. Here we present a brief discussion of models where the real component of the scalar $\varphi$ field is light. The strong $SU(N_b)$ is demoted to a global flavor symmetry, and $m_\varphi^2 < 0$ such that the scalar field develops a non-zero VEV via the Higgs mechanism. For $\lambda_{\varphi h} = 0$, we have
\begin{equation}
  \langle \varphi \rangle \equiv \frac{v_\varphi}{\sqrt{2}} = \sqrt{ - \frac{m_{\varphi}^2}{2 \lambda_\varphi} } \, .
\end{equation}
Expanding about the minimum, $\varphi = (v_\varphi + \phi_R + i \phi_I) / \sqrt{2}$, the imaginary component $\phi_I$ is eaten by the $Z^\prime$ gauge boson and the remaining scalar degree of freedom has mass
\begin{equation}
  m_{\phi_R} = \sqrt{\frac{\lambda_\varphi v_\varphi^2}{2}} = \sqrt{- \frac{m_\varphi^2}{2}} \, .
\end{equation}
We set the real scalar $\phi_R$ to be at the benchmark mass of 1.5 TeV, and require a cross section $\sigma(pp \rightarrow \phi_R \rightarrow \gamma \gamma) \simeq 0.1$ fb. The gluon fusion production cross section is given by
\begin{equation}
  \hat{\sigma}(gg \rightarrow \phi_R) \approx \frac{N_b^2 \alpha_s^2}{72 \sqrt{2} \pi} \frac{m_{\phi_R}^2}{v_\varphi^2} \, \delta ( \hat{s} - m_{\phi_R}^2 ) \, ,
\end{equation}
and the branching fraction to two photons is approximately Br$(\phi_R \rightarrow \gamma \gamma) \approx 2 \alpha^2 / 3 \alpha_s^2 = 4 \times 10^{-3}$. Using the {MSTW2008 NLO PDFs~\cite{Martin:2009iq} and including a $K$-factor of 2.5, we find the diphoton production cross section to be
\begin{equation}
  \sigma(p p \rightarrow \phi_R \rightarrow \gamma \gamma) \approx 0.73 \text{ fb} \times \left ( \frac{1 \text{ TeV}}{v_\varphi} \right )^2 \left ( \frac{N_b}{4} \right )^2 \, .
\end{equation}
Requiring a phenomenologically-motivated cross section of 0.1 fb, we anticipate $v_\varphi \sim 2-3$~TeV for $N_b = 3-5$.
    
The bare fermion masses are given by $m_{\psi_{1,2}} = y_{1,2}\,v_\varphi / \sqrt{2}$. The strongest bounds come from the QCD triplet masses, which are constrained to be $m_{\psi^{\rm T}} \gtrsim 1$ TeV \cite{Khachatryan:2015gza} for $\psi^{\rm T}$ mixed with the bottom quark. For $N_b = 4$ and $y_1^{\rm T,D} = y_2^{\rm T,D} \equiv y_{\rm T,D}$, this requires a Yukawa coupling $y_{\rm T} \gtrsim 0.5$. Requiring these relatively large Yukawa couplings at $m_\psi$ leads to Landau poles below the GUT scale in models with our matter content. For $N_b = 4$, the coupling becomes non-perturbative ($y > 4 \pi / \sqrt{2 N_b}$) at ${\cal O}(10^7 \text{ GeV})$. Fig.~\ref{fig:yukawa_landau} shows the Yukawa coupling running for $N_b = 3, 4, 5$.

\begin{figure}[ht]
  \begin{center}
    \includegraphics[scale=0.6]{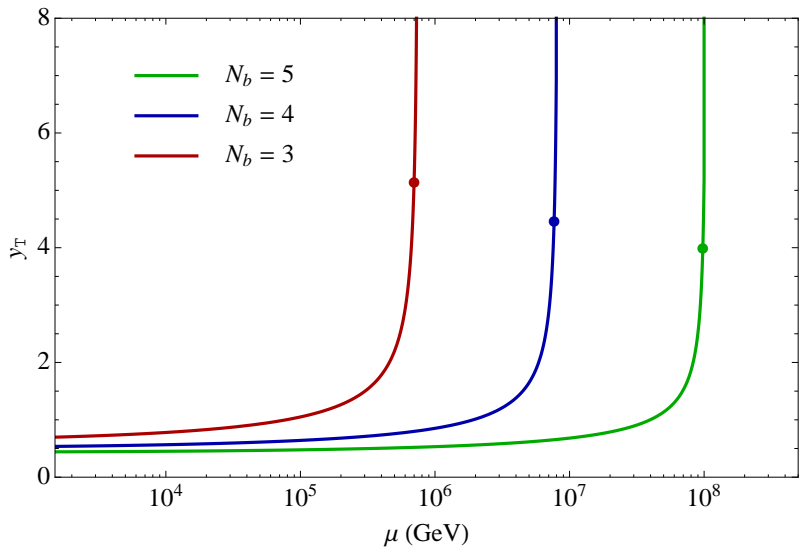}
    \caption{One loop runnings of $y_{\rm T}$ for the elementary scalar model, for $y^T(1.5~\mbox{TeV}) = y^D(1.5~\mbox{TeV})$. The dots show the approximate value at which the Yukawa couplings become non-perturbative.}
    \label{fig:yukawa_landau}
  \end{center}
\end{figure}
    
Furthermore, the tree level coupling $\phi_R Z^\prime Z^\prime$ will dominate the decay processes unless it is kinematically forbidden, which provides the additional constraint $m_{Z^\prime} > m_{\phi_R} / 2$. The $Z^\prime$ mass is given by
\begin{equation}
  m_{Z^\prime} = g^\prime \left | q_1 - q_2 \right | v_\varphi = g^\prime \left | q_1 - q_2 \right | \sqrt{ - \frac{m_\varphi^2}{\lambda_\varphi} } \, .
\end{equation}
For $N_b = 4$, this suggests values of the $U(1)'$ gauge coupling of $g' \left | q_1 - q_2 \right | \gtrsim 0.28$ at $m_{\phi_R}$. 

\section{$U(1)'$ Gauge Coupling Running}
\label{sec:gprime-running}
The one-loop beta function for the $U(1)^\prime$ gauge coupling running is
\beqa
\frac{dg^\prime}{d\log{\mu}} \equiv \beta(g^\prime) = \frac{g'^{3}}{16\pi^2} \left[ \frac{1}{3} \sum_{s} q_{s}^{2} + \frac{2}{3}\sum_{f} q_{f}^{2}\right] \, ,
\eeqa
for complex scalars $s$ and Weyl fermions $f$. For the matter content given in Table \ref{tab:fieldcontent}, we have
\beqa
\beta(g^\prime) = \frac{g'^{3}}{16\pi^2} \left[ \frac{1}{3} (q_{1} - q_{2})^{2} + 5\,N_{b}\, \frac{2}{3}(2\,q_{1}^{2} + 2\,q_{2}^{2})\right] \, .
\eeqa
Solving this equation, we find
\beqa
\frac{1}{g^{\prime 2}(M_{\rm GUT} )} = \frac{1}{g^{\prime 2}(\mu)} - \frac{1}{8\pi^{2}}\left[ \frac{1}{3} (q_{1} - q_{2})^{2} +\frac{20N_{b}}{3}(q_{1}^{2} + q_{2}^{2})\right] \log\left( \frac{M_{\rm GUT} }{\mu}  \right) \,.
\eeqa
To have the Landau pole occur at or below $M_{\rm GUT} \approx 3 \times 10^{16}$~GeV, we need to have (for $q_{1} = 1$ and $q_{2} = 0$)
\beqa
g^\prime(1~{\rm TeV}) \lesssim \frac{2.76}{\sqrt{1+20\,N_{b}}} \,, 
\eeqa
which is
\beqa
g^\prime(1~{\rm TeV}) \lesssim (0.35, 0.31, 0.28)\qquad \qquad \mbox{for}\,\quad N_b=(3, 4, 5) \,.
\eeqa
%

\section{Doublet and Triplet Yukawa Coupling Running}
\label{app:yd-over-yt}
The beta functions for the doublet and triplet Yukawa couplings are 
\begin{equation}
\mu^{2} \frac{d y_{\rm T}}{d\mu^{2}} = \frac{y_{\rm T}}{16\pi^{2}} \left[ \frac{3}{4} y_{\rm T}^{2} + \frac{N_{f}}{2}(3y_{\rm T}^{2}+2y_{\rm D}^{2}) - 3C^{\rm F}_{3}\, g_{s}^{2}\right] \, ,
\end{equation}
\begin{equation}
\mu^{2} \frac{d y_{\rm D}}{d\mu^{2}} = \frac{y_{\rm D}}{16\pi^{2}} \left[ \frac{3}{4} y_{\rm D}^{2} + \frac{N_{f}}{2}(3y_{\rm T}^{2}+2y_{\rm D}^{2}) - 3C^{\rm F}_{2}\, g_{2}^{2}\right] \, ,
\end{equation}
where $N_f= 2 N_b$; $C^{\rm F}_3 = 4/3$ is the quadratic Casimir of $SU(3)_c$; $C^{\rm F}_2 = 3/2$ is the quadratic Casimir of $SU(2)_W$; the sub-leading gauge interactions are neglected. We impose the GUT scale boundary condition that $y_{\rm D}(\Lambda_{\rm GUT}) = y_{\rm T}(\Lambda_{\rm GUT}) = y_{0}$.  However, the running of the ratio $\mathcal{R}_{y} \equiv y_{\rm D}/y_{\rm T}$ is nearly independent of the value of $y_{0}$. To see this, we re-write the differential equations for $y_{\rm T}$ and $y_{\rm D}$ in terms of $\mathcal{R}_{y}$ and $y_{\rm T}$ as follows
\begin{equation}
\mu^{2} \frac{d \mathcal{R}_{y}}{d\mu^{2}} = \frac{\mathcal{R}_{y}}{16\pi^{2}} \left[ \frac{3}{4} \, y_{\rm T}^{2} (\mathcal{R}_{y}^{2} - 1) + 3(C^{\rm F}_{3} \, g_{s}^{2} - C^{\rm F}_{2} \, g_{2}^{2})\right] \, ,
\end{equation}
\begin{equation}
\mu^{2} \frac{d y_{\rm T}}{d\mu^{2}} = \frac{y_{\rm T}}{16\pi^{2}} \left[ \frac{3}{4} y_{\rm T}^{2} + \frac{N_{f}}{2}\, y_{\rm T}^{2}(3+2\mathcal{R}_{y}^{2}) - 3C^{\rm F}_{3}\, g_{s}^{2}\right] \, ,
\end{equation}
where the new GUT scale boundary conditions are $\mathcal{R}_{y} (\Lambda_{\rm GUT}) = y_{\rm D}(\Lambda_{\rm GUT}) / y_{\rm T}(\Lambda_{\rm GUT}) = 1$ and $y_{\rm T}(\Lambda_{\rm GUT}) = y_{0}$. We see that the initial running for $\mathcal{R}_{y}$ does not depend on the value of $y_{0}$; it is determined only by the GUT scale gauge couplings.
\begin{figure}[h]
\includegraphics[scale=0.6]{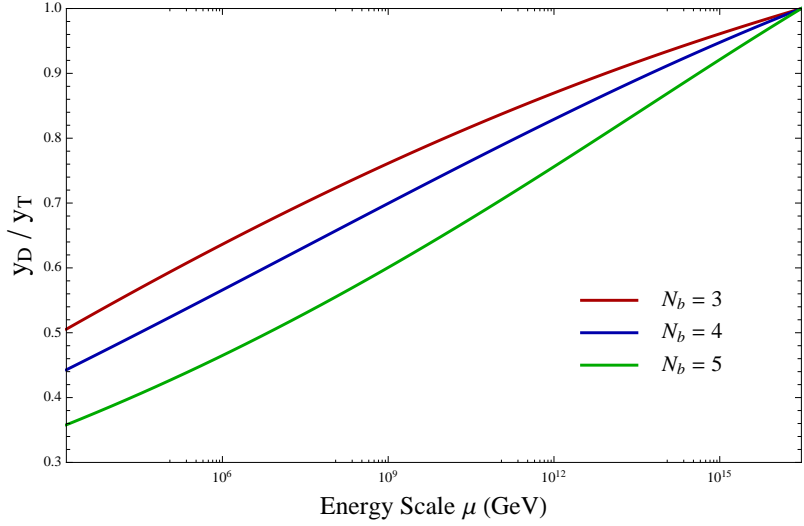}
\centering
\caption{The running of $y_{\rm D}/y_{\rm T}$ as a function of energy scale $\mu$. Boundary condition is chosen to be $y_{\rm D}(\Lambda_{\rm GUT})/y_{\rm T}(\Lambda_{\rm GUT}) = 1$ with $\Lambda_{\rm GUT} \approx 3\times 10^{16}$ GeV. }
\label{fig:test}
\end{figure}
Numerical integration of these equations is shown in Fig.~\ref{fig:test}. The numerical values for $y_{\rm D}/y_{\rm T}$ at $\mu =1.5$ TeV are shown in Table \ref{tab:RyTeV}.
\begin{table}[htb!]
\renewcommand{\arraystretch}{2.2}
  \centering
  \begin{tabular}{|c|c|c|c|}
    \hline \hline
 $N_{b}$ & 3 & 4 & 5 \\ \hline
$\frac{y_{\rm D}}{y_{\rm T}}(\mu = 1.5 \, {\rm TeV})$ & 0.51 & 0.44 & 0.36 \\ \hline
    \hline
  \end{tabular}
  \caption{ Numerical values for the ratios of the doublet and triplet Yukawa couplings at $\mu = 1.5$~TeV generated from a GUT scale boundary condition of $y_{\rm D}(\Lambda_{\rm GUT}) / y_{\rm T}(\Lambda_{\rm GUT}) = 1$.}  \label{tab:RyTeV}
\end{table}
Additionally, to avoid a Landau pole at or before $\mu = \Lambda_{\rm GUT} \sim 3\times 10^{16}$ GeV, we require
\beqa
y_{\rm T} (\mu = 1.5 \, {\rm TeV}) <  (0.54, 0.50, 0.49)  \hspace{10mm}  (N_b = 3, 4, 5) \,.
\eeqa
%

\section{$Z'$ Kinetic Mixing and Decay}
\label{app:Zprime-diagonalize}
\subsection{Gauge Kinetic Mixing}
To leading order in the gauge kinetic mixing parameter $\chi$, the process of diagonalizing the field strengths, and then resulting gauge boson mass terms after all symmetry breaking yields the following linear combinations for the gauge bosons \cite{Babu:1997st}
\begin{equation}
\hat{A}_{\mu} = A_{\mu} - \chi \, c_W  \sin{\xi} \, Z_{\mu} -  \chi \, c_W\cos{\xi}\, Z'_{\mu} \, ,
\end{equation}
\begin{equation}
\hat{Z}_{\mu} = \left( \cos\xi + \chi \, s_W \sin\xi \right) Z_{\mu} + \left(\chi\, s_W \cos\xi - \sin\xi \right) Z'_{\mu} \, ,
\end{equation}
\begin{equation}
\hat{Z}'_{\mu} = \sin\xi \, Z_{\mu} + \cos\xi \, Z'_{\mu}\, ,
\end{equation}
where
\begin{equation}
\tan 2\xi = \frac{2\, \chi \, s_W}{ 1-(\hat{m}_{Z'}/\hat{m}_{Z})^{2}} \, .
\end{equation}

\subsection{Decay of the $Z'$ Gauge Boson to Standard Model Fermions}
After diagonalization of the gauge boson kinetic and mass terms, we find that the $Z'$ gauge boson has a milli-charge coupling to the SM neutral current, given as
\begin{equation}
\mathcal{L}_{N} = \frac{g_{2}}{\cos\theta_{W}}\left[\Theta_{ZZ'} \, J_{3}^{\mu} + (s_W\,c_W\,\Theta_{AZ'} - s_W^{2}\, \Theta_{ZZ'})\, J^{\mu}_{\rm EM} \right]  Z'_{\mu} \, ,
\end{equation}
with
\begin{equation}
\Theta_{AZ'} = - \chi \, c_W \cos\xi \, , \hspace{20mm} \Theta_{ZZ'} = \chi \, s_W\cos\xi - \sin\xi \, .
\end{equation}
Inserting the Standard Model fermion current, the part of the Lagrangian relevant for $Z'$ decay can be cast in a familiar form
\begin{equation}
\mathcal{L}_{N} \supset \frac{g_{2}}{2\,c_W} \bar{\psi} \gamma^{\mu} (\mathcal{E}_{V} -\mathcal{E}_{A} \gamma^{5}) \psi \, Z'_{\mu} \, ,
\end{equation}
with the vector and axial milli-couplings defined as
\begin{equation}
\mathcal{E}_{V} = (T^{3} - 2Qs^{2}_W) \, \Theta_{ZZ'} + 2 Q \, s_W\,c_W\Theta_{AZ'}\, , \hspace{20mm} \mathcal{E}_{A} = T^{3} \, \Theta_{ZZ'} \, .
\end{equation}
From this, we find that the $Z'$ decay width to Standard Model fermions is
\begin{equation}
\Gamma (Z' \rightarrow \bar{\psi}\psi) = \frac{g_{2}^{2}}{48\pi\,c_W^2} \left( \mathcal{E}_{V}^{2} + \mathcal{E}_{A}^{2}   \right) m_{Z'} \, .
\end{equation}
\subsection{Decay of the $Z^\prime$ Gauge Boson to $Zh$ and $W^{+}W^{-}$}
If the $Z'$ gauge boson is heavy enough, there are also decay channels for $Z' \rightarrow Z \, h$ and $Z' \rightarrow W^{+}W^{-}$. The decay to $Zh$ comes from the $hZZ$ vertex in the Standard Model, which has a coupling of $g_{hZZ} = m_{Z}^{2}/v_{\rm EW}$, where $v_{\rm EW} = 246$ GeV is the electroweak VEV. After diagonalization of the gauge boson kinetic and mass terms, we have the following term in the Lagrangian
\begin{equation}
\mathcal{L}_{h} \supset g_{hZZ} \, (2 \chi \, s_W \cos 2\xi -\sin 2\xi) \, h \, Z_{\mu}Z'^{\mu} \, 
\equiv g_{hZZ'} \, h \, Z_{\mu}Z'^{\mu} \, ,
\end{equation}
at ${\cal O}(\chi)$, which allows the decay $Z' \rightarrow Z \, h$. The decay width is \cite{Barger:1987xw}
\begin{equation}
\Gamma(Z' \rightarrow Z\, h) = \frac{g_{hZZ'}^{2}\, m_{Z'}}{192\pi} \sqrt{\lambda(1,r_{Z}^{2},r_{h}^{2})} \left[\lambda(1,r_{Z}^{2},r_{h}^{2}) + 12 \, r_{Z}^{2} \right]\, ,
\end{equation}
where $r_{Z} = m_{Z}/m_{Z'}$, $r_{h} = m_{h}/m_{Z'}$, and $\lambda(a,b,c) = a^{2} + b^{2} + c^{2} - 2ab - 2ac - 2bc$.
The $Z' \rightarrow W^{+}W^{-}$ decay is mediated by the $WWA$ and $WWZ$ vertices in the Standard Model, which have couplings of $g_{WWA} = e$ and $g_{WWZ} = e\cot\theta_{W}$. The $Z'$ boson will therefore couple to $W^{+}W^{-}$ with strength
\begin{equation}
g_{WWZ'} =  e\, \left[ \cot\theta_{W} \left(s_W\chi \cos\xi - \sin\xi \right) - c_W\chi \cos{\xi} \right] = - e \cot\theta_{W} \sin\xi \, ,
\end{equation}
which leads to a decay width of \cite{Leike:1998wr}
\begin{equation}
\Gamma(Z' \rightarrow W^{+}W^{-}) = \frac{g^{2}_{WWZ'}\, m_{Z'}}{192\pi} \left( \frac{m_{Z'}}{m_{Z}}\right)^{4} (1-4 \, r_{W}^{2})^{3/2} \left( 1+20 \, r_{W}^{2} + 12 \, r_{W}^{4}\right) \, ,
\end{equation}
where $r_{W} = m_{W}/m_{Z'}$.


\begin{thebibliography}{10}

\bibitem{Hill:2002ap}
C.~T. Hill and E.~H. Simmons, {\it {Strong dynamics and electroweak symmetry
  breaking}},  {\em Phys. Rept.} {\bf 381} (2003) 235--402,
  [\href{http://arxiv.org/abs/hep-ph/0203079}{{\tt hep-ph/0203079}}]. [Erratum:
  Phys. Rept.390,553(2004)].

\bibitem{He:2001tp}
H.-J. He, N.~Polonsky, and S.-f. Su, {\it {Extra families, Higgs spectrum and
  oblique corrections}},  {\em Phys. Rev.} {\bf D64} (2001) 053004,
  [\href{http://arxiv.org/abs/hep-ph/0102144}{{\tt hep-ph/0102144}}].

\bibitem{Novikov:2001md}
V.~A. Novikov, L.~B. Okun, A.~N. Rozanov, and M.~I. Vysotsky, {\it {Extra
  generations and discrepancies of electroweak precision data}},  {\em Phys.
  Lett.} {\bf B529} (2002) 111--116,
  [\href{http://arxiv.org/abs/hep-ph/0111028}{{\tt hep-ph/0111028}}].

\bibitem{Kribs:2007nz}
G.~D. Kribs, T.~Plehn, M.~Spannowsky, and T.~M.~P. Tait, {\it {Four generations
  and Higgs physics}},  {\em Phys. Rev.} {\bf D76} (2007) 075016,
  [\href{http://arxiv.org/abs/0706.3718}{{\tt arXiv:0706.3718}}].

\bibitem{Erler:2010sk}
J.~Erler and P.~Langacker, {\it {Precision Constraints on Extra Fermion
  Generations}},  {\em Phys. Rev. Lett.} {\bf 105} (2010) 031801,
  [\href{http://arxiv.org/abs/1003.3211}{{\tt arXiv:1003.3211}}].

\bibitem{Eberhardt:2012gv}
O.~Eberhardt, G.~Herbert, H.~Lacker, A.~Lenz, A.~Menzel, U.~Nierste, and
  M.~Wiebusch, {\it {Impact of a Higgs boson at a mass of 126 GeV on the
  standard model with three and four fermion generations}},  {\em Phys. Rev.
  Lett.} {\bf 109} (2012) 241802, [\href{http://arxiv.org/abs/1209.1101}{{\tt
  arXiv:1209.1101}}].

\bibitem{Iwasaki:2003de}
Y.~Iwasaki, K.~Kanaya, S.~Kaya, S.~Sakai, and T.~Yoshie, {\it {Phase structure
  of lattice QCD for general number of flavors}},  {\em Phys. Rev.} {\bf D69}
  (2004) 014507, [\href{http://arxiv.org/abs/hep-lat/0309159}{{\tt
  hep-lat/0309159}}].

\bibitem{Ryttov:2007sr}
T.~A. Ryttov and F.~Sannino, {\it {Conformal Windows of SU(N) Gauge Theories,
  Higher Dimensional Representations and The Size of The Unparticle World}},
  {\em Phys. Rev.} {\bf D76} (2007) 105004,
  [\href{http://arxiv.org/abs/0707.3166}{{\tt arXiv:0707.3166}}].

\bibitem{Appelquist:2012nz}
T.~Appelquist et~al., {\it {Approaching Conformality with Ten Flavors}},
  \href{http://arxiv.org/abs/1204.6000}{{\tt arXiv:1204.6000}}.

\bibitem{Deuzeman:2013kma}
A.~Deuzeman, M.~P. Lombardo, K.~Miura, T.~Nunes~da Silva, and E.~Pallante, {\it
  {Phases of many flavors QCD : Lattice results}},  {\em PoS} {\bf
  ConfinementX} (2012) 274, [\href{http://arxiv.org/abs/1304.3245}{{\tt
  arXiv:1304.3245}}].

\bibitem{Lombardo:2014pda}
M.~P. Lombardo, K.~Miura, T.~J. Nunes~da Silva, and E.~Pallante, {\it {On the
  particle spectrum and the conformal window}},  {\em JHEP} {\bf 12} (2014)
  183, [\href{http://arxiv.org/abs/1410.0298}{{\tt arXiv:1410.0298}}].

\bibitem{Fodor:2016zil}
Z.~Fodor, K.~Holland, J.~Kuti, S.~Mondal, D.~Nogradi, and C.~H. Wong, {\it
  {Fate of the conformal fixed point with twelve massless fermions and SU(3)
  gauge group}},  \href{http://arxiv.org/abs/1607.06121}{{\tt
  arXiv:1607.06121}}.

\bibitem{Peskin:1980gc}
M.~E. Peskin, {\it {The Alignment of the Vacuum in Theories of Technicolor}},
  {\em Nucl. Phys.} {\bf B175} (1980) 197--233.

\bibitem{Bai:2010qg}
Y.~Bai and R.~J. Hill, {\it {Weakly Interacting Stable Pions}},  {\em Phys.
  Rev.} {\bf D82} (2010) 111701, [\href{http://arxiv.org/abs/1005.0008}{{\tt
  arXiv:1005.0008}}].

\bibitem{Antipin:2015xia}
O.~Antipin, M.~Redi, A.~Strumia, and E.~Vigiani, {\it {Accidental Composite
  Dark Matter}},  {\em JHEP} {\bf 07} (2015) 039,
  [\href{http://arxiv.org/abs/1503.08749}{{\tt arXiv:1503.08749}}].

\bibitem{Bai:2015nbs}
Y.~Bai, J.~Berger, and R.~Lu, {\it {A 750 GeV Dark Pion: Cousin of a Dark
  G-parity-odd WIMP}},  {\em Phys. Rev.} {\bf D93} (2016) 076009,
  [\href{http://arxiv.org/abs/1512.05779}{{\tt arXiv:1512.05779}}].

\bibitem{Redi:2016kip}
M.~Redi, A.~Strumia, A.~Tesi, and E.~Vigiani, {\it {Di-photon resonance and
  Dark Matter as heavy pions}},  \href{http://arxiv.org/abs/1602.07297}{{\tt
  arXiv:1602.07297}}.
  
\bibitem{Khachatryan:2016yec}
{\bf CMS} Collaboration, V.~Khachatryan et~al., {\it {Search for high-mass
  diphoton resonances in proton-proton collisions at 13 TeV and combination
  with 8 TeV search}},  \href{http://arxiv.org/abs/1609.02507}{{\tt
  arXiv:1609.02507}}.
  
\bibitem{ATLAS:2016eeo}
{\bf ATLAS} Collaboration, T.~A. collaboration, {\it {Search for scalar
  diphoton resonances with 15.4~fb$^{-1}$ of data collected at $\sqrt{s}$=13
  TeV in 2015 and 2016 with the ATLAS detector}}, .

\bibitem{Holdom:1985ag}
B.~Holdom, {\it {Two U(1)'s and Epsilon Charge Shifts}},  {\em Phys. Lett.}
  {\bf B166} (1986) 196--198.

\bibitem{Hook:2010tw}
A.~Hook, E.~Izaguirre, and J.~G. Wacker, {\it {Model Independent Bounds on
  Kinetic Mixing}},  {\em Adv. High Energy Phys.} {\bf 2011} (2011) 859762,
  [\href{http://arxiv.org/abs/1006.0973}{{\tt arXiv:1006.0973}}].

\bibitem{Pospelov:2008zw}
M.~Pospelov, {\it {Secluded U(1) below the weak scale}},  {\em Phys. Rev.} {\bf
  D80} (2009) 095002, [\href{http://arxiv.org/abs/0811.1030}{{\tt
  arXiv:0811.1030}}].

\bibitem{Aubert:2009af}
{\bf BaBar} Collaboration, B.~Aubert et~al., {\it {Search for a Narrow
  Resonance in e+e- to Four Lepton Final States}},  in {\em {Proceedings, 24th
  International Symposium on Lepton-Photon Interactions at High Energy
  (LP09)}}, 2009.
\newblock \href{http://arxiv.org/abs/0908.2821}{{\tt arXiv:0908.2821}}.

\bibitem{LEP:2003aa}
{\bf SLD Electroweak Group, SLD Heavy Flavor Group, DELPHI, LEP, ALEPH, OPAL,
  LEP Electroweak Working Group, L3} Collaboration, t.~S. Electroweak, {\it {A
  Combination of preliminary electroweak measurements and constraints on the
  standard model}},  \href{http://arxiv.org/abs/hep-ex/0312023}{{\tt
  hep-ex/0312023}}.

\bibitem{Martin:2009iq}
A.~D. Martin, W.~J. Stirling, R.~S. Thorne, and G.~Watt, {\it {Parton
  distributions for the LHC}},  {\em Eur. Phys. J.} {\bf C63} (2009) 189--285,
  [\href{http://arxiv.org/abs/0901.0002}{{\tt arXiv:0901.0002}}].

\bibitem{Catani:2003zt}
S.~Catani, D.~de~Florian, M.~Grazzini, and P.~Nason, {\it {Soft gluon
  resummation for Higgs boson production at hadron colliders}},  {\em JHEP}
  {\bf 07} (2003) 028, [\href{http://arxiv.org/abs/hep-ph/0306211}{{\tt
  hep-ph/0306211}}].

\bibitem{Idilbi:2009cc}
A.~Idilbi, C.~Kim, and T.~Mehen, {\it {Factorization and resummation for single
  color-octet scalar production at the LHC}},  {\em Phys. Rev.} {\bf D79}
  (2009) 114016, [\href{http://arxiv.org/abs/0903.3668}{{\tt
  arXiv:0903.3668}}].

\bibitem{Bai:2016czm}
Y.~Bai, V.~Barger, and J.~Berger, {\it {Color-octet Companions of a 750 GeV
  Heavy Pion}},  \href{http://arxiv.org/abs/1604.07835}{{\tt
  arXiv:1604.07835}}.

\bibitem{Bai:2010dj}
Y.~Bai and B.~A. Dobrescu, {\it {Heavy octets and Tevatron signals with three
  or four b jets}},  {\em JHEP} {\bf 07} (2011) 100,
  [\href{http://arxiv.org/abs/1012.5814}{{\tt arXiv:1012.5814}}].

\bibitem{Chivukula:2016zbe}
R.~S. Chivukula, A.~Farzinnia, K.~Mohan, and E.~H. Simmons, {\it {Diphoton
  Resonances in the Renormalizable Coloron Model}},
  \href{http://arxiv.org/abs/1604.02157}{{\tt arXiv:1604.02157}}.

\bibitem{Chivukula:2013xka}
R.~S. Chivukula, A.~Farzinnia, J.~Ren, and E.~H. Simmons, {\it {Constraints on
  the Scalar Sector of the Renormalizable Coloron Model}},  {\em Phys. Rev.}
  {\bf D88} (2013), no.~7 075020, [\href{http://arxiv.org/abs/1307.1064}{{\tt
  arXiv:1307.1064}}]. [Erratum: Phys. Rev.D89,no.5,059905(2014)].

\bibitem{Buttazzo:2016kid}
D.~Buttazzo, A.~Greljo, G.~Isidori, and D.~Marzocca, {\it {Toward a coherent
  solution of diphoton and flavor anomalies}},
  \href{http://arxiv.org/abs/1604.03940}{{\tt arXiv:1604.03940}}.

\bibitem{Harigaya:2016pnu}
K.~Harigaya and Y.~Nomura, {\it {A Composite Model for the 750 GeV Diphoton
  Excess}},  {\em JHEP} {\bf 03} (2016) 091,
  [\href{http://arxiv.org/abs/1602.01092}{{\tt arXiv:1602.01092}}].

\bibitem{Khachatryan:2015jha}
{\bf CMS} Collaboration, V.~Khachatryan et~al., {\it {Search for Decays of
  Stopped Long-Lived Particles Produced in Proton-Proton Collisions at
  $\sqrt{s}= 8\,\text {TeV} $}},  {\em Eur. Phys. J.} {\bf C75} (2015), no.~4
  151, [\href{http://arxiv.org/abs/1501.05603}{{\tt arXiv:1501.05603}}].

\bibitem{Chatrchyan:2013oca}
{\bf CMS} Collaboration, S.~Chatrchyan et~al., {\it {Searches for long-lived
  charged particles in pp collisions at $\sqrt{s}$=7 and 8 TeV}},  {\em JHEP}
  {\bf 07} (2013) 122, [\href{http://arxiv.org/abs/1305.0491}{{\tt
  arXiv:1305.0491}}].

\bibitem{Khachatryan:2015gza}
{\bf CMS} Collaboration, V.~Khachatryan et~al., {\it {Search for pair-produced
  vector-like B quarks in proton-proton collisions at $\sqrt{s}$ = 8 TeV}},
  \href{http://arxiv.org/abs/1507.07129}{{\tt arXiv:1507.07129}}.

\bibitem{Babu:1997st}
K.~S. Babu, C.~F. Kolda, and J.~March-Russell, {\it {Implications of
  generalized Z - Z-prime mixing}},  {\em Phys. Rev.} {\bf D57} (1998)
  6788--6792, [\href{http://arxiv.org/abs/hep-ph/9710441}{{\tt
  hep-ph/9710441}}].

\bibitem{Barger:1987xw}
V.~D. Barger and K.~Whisnant, {\it {Heavy $Z$ Boson Decays to Two Bosons in
  $E(6)$ Superstring Models}},  {\em Phys. Rev.} {\bf D36} (1987) 3429.

\bibitem{Leike:1998wr}
A.~Leike, {\it {The Phenomenology of extra neutral gauge bosons}},  {\em Phys.
  Rept.} {\bf 317} (1999) 143--250,
  [\href{http://arxiv.org/abs/hep-ph/9805494}{{\tt hep-ph/9805494}}].

\end{thebibliography}
\providecommand{\href}[2]{#2}\begingroup\raggedright\endgroup

 \end{document}